\documentclass[mathpazo]{cicp}

\usepackage{subfigure}
\usepackage{amsmath,amssymb,amsfonts,amsthm}
\usepackage{bm}
\usepackage{graphicx}
\usepackage{newclude}
\usepackage{url}
\usepackage{color}
\usepackage{multirow}
\usepackage[mathscr]{euscript}
\usepackage{mdframed,booktabs,rotating}
\usepackage[suffix=]{epstopdf}
\DeclareGraphicsExtensions{.png,.pdf}
\usepackage[title,toc,titletoc,page]{appendix}
\graphicspath{{pic/}}
\usepackage[numbers,sort&compress]{natbib} % CiCP need

\usepackage[linesnumbered,ruled,vlined]{algorithm2e}
\usepackage{epstopdf}
\begin{document}
%%%%% title : short title may not be used but TITLE is required.
% \title{TITLE}
% \title[short title]{TITLE}
\title{A Robust and Efficient Multi-physics Numerical System for Intensive Blast Wave Propagation in Complex Environments}

\author[Huang M S et.~al.]{Minsheng Huang\affil{1},
       Pan Wang\affil{2}, Chengbao Yao\affil{2}\comma\corrauth, Lidong Cheng\affil{3}\comma\affil{*}, Wenjun Ying\affil{1}\comma\affil{4}\comma\affil{*}}
\address{\affilnum{1}\ School of Mathematical Sciences, 
Shanghai Jiao Tong University, 
Shanghai 200240, P.R. China. \\
\affilnum{2}\ Northwest Institute of Nuclear Technology, 
Xi'an 710038, P.R. China. \\
\affilnum{3}\ School of Aeronautics and Astronautics, 
Shanghai Jiao Tong University, 
Shanghai 200240, P.R. China. \\
\affilnum{4}\ MOE-LSC and Institute of Natural Sciences, 
Shanghai Jiao Tong University, 
Shanghai 200240, P.R. China.}
\emails{{\tt yaocheng@pku.edu.cn} (C.~Yao), {\tt critters@sjtu.edu.cn} (L.~Cheng), {\tt wying@sjtu.edu.cn} (W.~Ying)}
%%%%% Begin Abstract %%%%%%%%%%%
\begin{abstract}
We establish a high-resolution, high-performance, and high-confidence compressible multiphysics system in a Cartesian grid with irregular boundary topologies to simulate intensive blast waves propagating in large-scale and extremely complex environments. The multiphysics system is modeled by a multi-component model solved using a generalized Godunov method and a classical material point method in a combination of Lagrangian particles and a rigid material model. An artificial neural network equation of state (EOS) is proposed based on experimental data to simulate the intensive explosion products and real gas under extreme pressure and temperature. To improve computational accuracy and efficiency, a deepMTBVD reconstruction scheme of our previous work is extended to the multiphysics system. With the aid of high-performance parallel computation, several large-scale blast wave applications, such as blast wave propagating in a local and entire urban city, are simulated in a reasonable time period, which can validate numerical schemes and lead to more practical engineering applications. 
\end{abstract}
%%%%% end %%%%%%%%%%%

%%%%% AMS/PACs/Keywords %%%%%%%%%%%
%\pac{}

%%%% maketitle %%%%%
\maketitle

%%%% Start %%%%%%
\section{Introduction}
Intensive blast waves are an essential area of research in the fields of physics, engineering, and military science. In recent decades, there has been rapidly growing interest in understanding the behavior of blast waves in complex environments, especially in urban and city areas \cite{wang2017urban,ratcliff2023,valsamos2021urban}. The study of blast waves in complex environments has made significant progress with advances in experimental, theoretical, and numerical methods. 
The studies have provided valuable insights into the dynamic behavior of blast waves in urban and city areas. The findings have significant implications for guiding the design and construction of resilient infrastructures. Furthermore, they contribute to the refinement of military strategies, as well as the mitigation of damage from explosive events \cite{jimenez2010,denny2024}.

% These studies have provided valuable insights into the behavior of blast waves in urban and city areas and have important implications for the design and construction of structures, the development of military strategies, and the mitigation of damage from explosive events.
Theoretical studies typically utilize analytical equations of shock relations and empirical formulas based on correlations from experimental data. While this approach is efficient in ideal conditions, it may be prone to inaccuracies in complex environments \cite{Swisdak1994, Britt2001, Hyde1992}. Experimental studies on blast waves in complex environments have been conducted using various testing methods, such as explosive detonations and shock tubes \cite{shi2023urban, Remennikov2003experi}. Researchers have investigated the effects of obstacles, such as buildings and walls, on the propagation of blast waves and the resulting damage to structures and materials. These experiments have provided valuable data for development of numerical models and validation of theoretical predictions. For more details, please refer to \cite{ratcliff2023, Drazin2018, Remennikov2003, Smith2006, Hao2016}.

Due to the great development of computational science, numerical simulations have been widely applied to supplement experimental and theoretical studies of blast waves in complex environments. 
The numerical methods are adopted in simulations, including finite difference (FD), finite volume method (FVM), and finite element methods (FEM), to solve the governing equations describing the blast waves in complex urban geometries.
% These simulations use numerical methods, such as finite difference (FD), finite volume method (FVM), and finite element methods (FEM), to solve the equations governing the behavior of blast waves in complex geometries. 
Through numerical simulation, extensive research has been carried out on a wide range of blast wave phenomena, including the propagation of blast waves in confined spaces, the effects of obstacles on blast wave propagation, and the interaction of blast waves with materials and structures.
A variety of commercial software and open source codes, such as LS-DYNA, AUTODYN, SHAMRC, ALE3D, blastFoam, ECOGEN, etc \cite{birnbaum1987autodyna, shamrc, noble_ale3d_2017, blastfoam, schmid2020ecogen}, can be applied to simulate the blast waves of chemical high explosives or gas detonation in a small and moderate scale of space, but lacks the ability to solve the large-scale problem with accurate equations of state under extremely thermodynamic conditions, especially in nuclear explosion applications. Representative works can be found in \cite{Hao2016, Drazin2018, Fu2022, Chen2018, Guo2016}.

The simulation of intense blast waves remains extremely challenging since it evolves the multi-material interactions, highly nonlinear EOS, multi-scale effects (especially in nuclear explosion cases), and complex topologies of the physical boundaries. Due to the difficulties mentioned above, the numerical solver should be established carefully to maintain robust and efficient behavior. In multiphase flow simulations, not only do the densities and pressures vary largely, but the constitutive relations also differ distinctly across the material interface. Therefore, the material interface between distinct fluids is critical in the modeling of multiphase flows. 
%\hms{Typically, there are two dominant types of numerical methods for multiphase flow simulation: the sharp interface method and the diffuse interface method. In the sharp interface method, the interface is assumed to be the sharp contact discontinuity and different fluids are immiscible. Several Eulerian approaches, such as volume of fluid (VOF) method \cite{Scardovelli1999, Noh1976}, level set method \cite{Sethian2001, Sussman1994}, moment of fluid (MOF) method \cite{Ahn2007, Dyadechko2008, Anbarlooei2009} and front-tracking method \cite{Glimm1998, Tryggvason2001} are used extensively to capture the interface. In the diffuse interface method, the physical models and governing equations are derived from phase field theory. The material interface is represented by a thin diffuse layer, and the flow properties, such as density and viscosity, change smoothly in the form of hyperbolic tangent functions. The algorithms of diffuse interface method, which have attained widespread applications in dealing with complex interface deformation problems \cite{Zein2010modeling, Saurel2018Diffuse}, are broadly classified into four major types: the seven-equation model \cite{Baer1986two, Sainsaulieu1995finite}, the six-equation model \cite{Saurel2009simple, saurel2008phase, Pelanti2014mixture, Pelanti2019numerical}, the five-equation model \cite{Kapila2001two, Allaire2002five, Garrick2017finite, Zhang2020simple} and the four-equation model \cite{Abgrall1996how, Johnsen2012Preventing, Movahed2013solution}. }
The mathematical models for multiphase flows, widely used to address complex interface deformation problems \cite{Zein2010modeling, Saurel2018Diffuse}, can be broadly classified into four major types: the seven-equation model \cite{Baer1986two, Sainsaulieu1995finite}, the six-equation model \cite{Saurel2009simple, saurel2008phase, Pelanti2014mixture, Pelanti2019numerical}, the five-equation model \cite{Kapila2001two, Allaire2002five, Garrick2017finite, Zhang2020simple} and the four-equation model \cite{Abgrall1996how, Johnsen2012Preventing, Movahed2013solution}.
Typically, there are two dominant types of numerical methods for multiphase flow simulation: the sharp interface method and the diffuse interface method. In the sharp interface method, the interface is assumed to be the sharp contact discontinuity, and different fluids are immiscible. In the diffuse interface method, the physical models and governing equations are derived from phase field theory. The material interface is represented by a thin, diffuse layer, and the flow properties, such as density and viscosity, change smoothly in the form of hyperbolic tangent functions.
Several Eulerian approaches, such as volume of fluid (VOF) method \cite{Scardovelli1999, Noh1976}, level set method \cite{Sethian2001, Sussman1994}, moment of fluid (MOF) method \cite{Ahn2007, Dyadechko2008, Anbarlooei2009} and front-tracking method \cite{Glimm1998, Tryggvason2001} are used extensively to capture the interface.

The sharp interface method is renowned for its accuracy in representing interface shapes. At the same time, it is more expensive due to the interface reconstruction and geometric advection processes within this method. The diffuse interface method makes no attempt to track the material interface but instead treats the flow as a mixture of two phases with an average mixture density. It is able to treat all the possible physics of multi-phase flow in the domain without special treatment, including the dynamical phase creation and interface creation, cavitation evolution, and collapse. The main drawback of the diffuse interface method is the excessive smearing of the fluid interfaces due to the numerical diffusion of the hyperbolic solver. To counteract this diffusion, a number of approaches have been developed, and they can be classified into two groups. In the first group, a high-order polynomial-based method, including the Essentially Non-Oscillatory (ENO) scheme \cite{harten1987eno, shu1988eno}, the Weighted Essentially Non-Oscillatory (WENO) scheme \cite{jiang1996WENO,shu2009reivew} is used. Methods in the second group use a series of correction terms, such as anti-diffusion, pseudo-time sharpening techniques \cite{Shukla2010interface, Shukla2014nonlinear, Nguyen2021numerical}, reconstruction-based interface sharpening approaches \cite{Tiwari2013diffuse, Nguyen2021numerical, Chiapolino2017sharpening} along with the governing equations to sharpen the profiles of volume fraction and density. Due to the presence of the Gibbs phenomenon \cite{Hewitt1979TheGP}, high-order polynomials are not optimal for reconstructing discontinuous flow fields. In contrast, sigmoid functions have been verified to perform better in such scenarios. The THINC (Tangent Hyperbolic Interface Capturing) scheme utilizes the specific hyperbolic tangent function for reconstruction \cite{XiaoHonma-13, XiaoIi-46, XieXiao-229}. Additionally, the BVD (Boundary Variation Diminishing) principle and schemes are proposed to take advantage of polynomial-based reconstruction in smooth regions and non-polynomial-based reconstruction in discontinuous regions. This approach allows the BVD scheme to accurately capture delicate flow structures with high fidelity, including resolving shock waves, reproducing multi-scale vortices, and capturing interfaces and contact discontinuities \cite{RN12, RN2, wakimura2022sym, DengShimizu-142}. However, BVD schemes require preparing all candidate reconstruction functions before selecting the final one within a given cell, suggesting that there is potential for improving computational efficiency \cite{huang2024struct}.

In this paper, we present a hybrid scheme that combines the multi-component diffuse interface method with the material point method (MPM) on Cartesian grids. This approach allows for the simulation of complex multiphase flows under extreme conditions, such as nuclear blast waves and high explosive shock waves, in large-scale urban environments. A seven-equation model with an arbitrary number of phases is adopted to simulate the multiphase compressible flows, and the general Godunov method is applied to calculate the numerical fluxes of the conservative and non-conservative terms. The MPM with the rigid solid constitutive model is adapted to simulate urban buildings and irregular ground. 
% Based on experimental data, an artificial neural network EOS is introduced to simulate the intensive explosion products and real gas under extreme pressure and temperature \cite{Li2023neural}. 
Due to the complexity of describing the thermodynamic variables within empirical or experimental formulas, we have established a neural network EOS to simulate the intensive explosion products and real gas under extreme pressure and temperature \cite{Li2023neural}. 
A new paradigm for constructing high-resolution hybrid schemes for compressible flows, in our previous work \cite{huang2024struct}, which generates training data based on MUSCL-THINC-BVD schemes \cite{deng2018high} for supervised learning and employs artificial neural networks to create an indicator that pre-selects the most suitable reconstruction scheme for each cell, is directly extended to the multiphase flow calculations. With the above numerical techniques and high-performance parallel computation, we establish a robust and efficient multi-physical numerical system to simulate the blast waves produced by various types of explosions in large-scale and complex urban environments. The numerical system is able to solve a classical blast wave application within a pressure range from $10^3$ to $10^{15}$ Pa, handle a total number of cells in a magnitude of tens of billions, and maintain a stable physical time of no less than several minutes. Using the simulated codes with tens of thousands of cores, we can accurately simulate the blast waves propagating in a local or entire urban city within a reasonable time period and obtain lots of valuable simulation data to assist the design and construction of structures and the development of military strategies. 

The rest of this paper is arranged as follows. In Section \ref{sec:model}, the governing equations of the multiphase fluid and solid, the constitutive models, and the equations of state in the application are introduced. The procedures of the multi-physical scheme and deepMTBVD reconstruction method are detailed in Section \ref{sec:scheme}. In Section \ref{sec:results}, several classical benchmark problems and typical air blast applications in complex urban environments are carried out to validate the accuracy and robustness of our schemes. Finally, a short conclusion is drawn in Section \ref{sec:conclusion}. 

%%% Local Variables: 
%%% mode: latex
%%% TeX-master: "article"
%%% End: 

% \section{Governing equations and constitutive equations} 
\section{Mathematical model}
\label{sec:model} 

\subsection{Governing equations of multiphase fluid}
% The governing equations for inviscid compressible multiphase fluid flows, in the absence of heat conduction and radiation, are a set of partial differential equations representing the balance of mass, momentum and energy for each phase. The governing equations, in the case of two phases, which can be written in a general form as \cite{Saurel2009simple}:
The governing equations for inviscid compressible multiphase fluid with different velocities and pressures in each phase can be written in a general form as \cite{Saurel2009simple}:
\begin{equation} 
\dfrac{\partial \bm{U}}{\partial t} + 
\nabla\cdot \bm F(\bm U) +
\bm B(\bm U)\cdot\nabla \alpha_1 = 
\bm H(\bm U),
\label{eq:cons} 
\end{equation}
where $\bm{U}$ is the state vector, $\bm{F}$ is the flux tensor, $\bm{B}$ and $\bm{H}$ is the no-conservative quantity and source term we will concretize subsequently. In this work, we ignore the heat conduction as well as radiation and consider a set of partial differential equations representing the balance of mass, momentum, and energy for each phase, which is described as \cite{Saurel1999multiphase}:
\[
\begin{split}
\bm{U}& = \begin{pmatrix}
\alpha_1 \\
\alpha_1 \rho_1 \\ 
\alpha_2 \rho_2 \\
\alpha_1 \rho_1 \bm u_1 \\ 
\alpha_2 \rho_2 \bm u_2 \\ 
\alpha_1 \rho_1 E_1 \\
\alpha_2 \rho_2 E_2
\end{pmatrix},~
\bm{F}(\bm{U})=
\begin{pmatrix}
0  \\
\alpha_1 \rho_1 \bm u_1  \\
\alpha_2 \rho_2 \bm u_2  \\
\alpha_1 (\rho_1 \bm u_1 \otimes \bm u_1 + p_1\mathbf I) \\
\alpha_2 (\rho_2 \bm u_2 \otimes \bm u_2 + p_2\mathbf I) \\
\alpha_1 (\rho_1 E_1+p_1) \bm u_1 \\
\alpha_2 (\rho_2 E_2+p_2) \bm u_2 
\end{pmatrix}, \\
\bm B(\bm U) & = 
\begin{pmatrix}
\bm u_I  \\
0  \\
0  \\
-p_I \\
 p_I \\
-p_I \bm u_I \\
 p_I \bm u_I
\end{pmatrix}, ~
\bm H(\bm U) = 
\begin{pmatrix}
\mu(p_1-p_2)  \\
0  \\
0  \\
\lambda (\bm u_2 - \bm u_1) \\
-\lambda (\bm u_2 - \bm u_1) \\
\mu p_I(p_2-p_1) + \lambda \bm u_I(\bm u_2 - \bm u_1) \\
-\mu p_I(p_2-p_1) - \lambda \bm u_I(\bm u_2 - \bm u_1) 
\end{pmatrix}.
\end{split}
\]
Here $\alpha_1, \alpha_2$ stand for the phasic volume fraction, $\rho_1, \rho_2$ denote the phasic density, $\bm u_1, \bm u_2$ indicate the phasic velocity, $E_1, E_2$ and $p_1, p_2$ stand for the phasic specific total energy and pressure, respectively. $\mu$ and $\lambda$ are the relaxation rates relating to the pressure and velocity relaxation procedure and will be discussed in the following section. $p_I$ and $\bm u_I$ are the interfacial pressure and velocity, which are defined from \cite{Saurel1999multiphase, saurel2001multi}:
\[
p_I=\alpha_1 p_1 + \alpha_2 p_2, \quad 
\bm u_I=\dfrac{\alpha_1\rho_1 \bm u_1+\alpha_2\rho_2\bm u_2}{\alpha_1\rho_1+\alpha_2\rho_2}.
\]

\subsection{Governing equations of solid}
Different from the multiphase fluid, the solid dynamic behavior is resolved within the Lagrangian framework, and the governing equations of mass, momentum and energy are formulated as 
\begin{equation}
\left\{
\begin{aligned}
\rho J &= \rho_0, \\
\rho \dot{\bm u} &= \nabla \cdot {\bm \sigma} + \rho \bm f, \\
 \dot{E} &= J {\bm \sigma} : \dot{\bm \varepsilon}
          = J({\mathbf S} : \dot{\bm \varepsilon}' - \dfrac{1}{3}p{\rm tr}(\dot{\varepsilon})).
\end{aligned}
\right.
\label{eq:solid}
\end{equation}
Here, $J$ stands for the determinant of the deformation gradient matrix ${\mathbf F}=\partial \bm x/\partial \bm X$, where $\bm x$ and $\bm X$ stand for the current and initial configuration, respectively. $\bm u$ is the velocity vector, $\rho_0$ is the initial density, and $\bm f$ is the body force density. $\bm \sigma$ is the Cauchy stress tensor, $\mathbf S$ is the deviatoric stress tensor, and $\mathbf{I}$ is the identity matrix, thus $\bm \sigma=-p\mathbf I + \mathbf S$. $\dot{\bm \varepsilon}$ is the Green strain rate tensor, $\dot{\bm \varepsilon}'$ is the deviatoric strain rate tensor, and ${\rm tr}(\dot{\varepsilon})$ is the trace of $\dot{\bm \varepsilon}$, which satisfies $\dot{\bm \varepsilon}=\dot{\bm \varepsilon}'+{\rm tr}(\dot{\varepsilon})\mathbf{I}$.

\subsection{Constitutive equations}
An additional constitutive equation relating to the thermodynamic variables is required to provide closure to the governing equations \eqref{eq:cons} and \eqref{eq:solid}. In the fluids phase, the constitutive equation is represented by an EOS with the form $p=p(\rho, e)$, in which $\rho, e$ denote the density and internal energy, respectively. In the phase of solids, the deformation is decomposed into the volumetric and shear deformation, and the Cauchy stress tensor is also divided into the hydrostatic pressure and deviatoric stress tensor, respectively. The hydrostatic pressure is still expressed by employing an equation of state-like fluids. In contrast, deviatoric stress is expressed by Hooke's law in the elastic region and the plastic flow rule in the plastic region. This paper focuses on the following constitutive equation to simulate our numerical applications. 

% \subsubsection*{\bf Ideal gas EOS}
$\bullet \textbf{ Ideal gas EOS}$

Most gases are often modeled using the ideal gas \cite{torobook}, which is expressed as follows:  
\begin{equation}
p = (\gamma-1) \rho e,
\label{eq:ideal}
\end{equation}
where $\gamma$ is the adiabatic exponent and $e$ denotes the internal energy.

% \subsubsection*{\bf JWL EOS}
$\bullet \textbf{ JWL EOS}$

The JWL EOS \cite{Lee1968jwl} is always applied to characterize various gaseous products of high explosives:
\begin{equation}
p = A_1\left(1-\frac{\omega\rho}{R_1\rho_0}\right)
\exp\left(-\frac{R_1\rho_0}{\rho}\right) +
A_2\left(1-\frac{\omega\rho}{R_2\rho_0}\right) \exp\left(-\frac{
    R_2\rho_0}{\rho}\right) + \omega\rho e,
\label{eq:jwl}
\end{equation}
where $A_1, A_2,\omega, R_1, R_2$ and $\rho_0$ are positive constants, which are determined empirically and experimentally in different products. Here we use the following values to describe the gaseous products of TNT \cite{Smith1999}: $A_1=3.712\times10^{11}\text{ Pa}, A_2=3.23\times10^9\text{ Pa}$, $\omega=0.30, R_1=4.15, R_2=0.95$ and $\rho_0=1630\text{ kg}/\text{m}^3$.

% \subsubsection*{\bf  Neural network nuclear products and real gas EOS}
$\bullet \textbf{ Neural network nuclear products and real gas EOS}$

Since the nuclear explosion is accompanied by intense heating, ionization, and other effects, the explosion products expand rapidly to a gas state, and the surrounding air is subject to intense heating and compression due to the combined effects of shock waves, thermal radiation, and other factors. The physical state is highly complex, while the traditional analytical EOS can not accurately describe its thermodynamic behaviors. In this paper, we choose the tabulated form of Saha EOS deduced from the experimental data and get the corresponding neural network EOS by training the tabulated data \cite{qiao2003}. The neural network EOS is expressed by the following formula, which has four hidden layers and ten neural nodes for each layer: 
\begin{equation}
p=p(\rho, e),
\label{eq:dnn}
\end{equation}
\[
\left\{ 
\begin{aligned}
a^{(0)} &= \left[\begin{array}{c} k_\rho\log{\rho}+c_{\rho} \\  k_e\log{e}+c_{e} \end{array}\right], \\
a^{(1)} &= \dfrac{2}{1+e^{-2\left(W^{(1)}a^{(0)}+b^{(1)} \right)}}-1, \\
a^{(2)} &= \dfrac{2}{1+e^{-2\left(W^{(2)}a^{(1)}+b^{(2)} \right)}}-1, \\
a^{(3)} &= \dfrac{2}{1+e^{-2\left(W^{(3)}a^{(2)}+b^{(3)} \right)}}-1, \\
a^{(4)} &= \dfrac{2}{1+e^{-2\left(W^{(4)}a^{(3)}+b^{(4)} \right)}}-1, \\
\bar{p} &= W^{(5)}a^{(4)}+b^{(5)}, \\
p &= e^{k_p\bar{p}+c_p}.
\end{aligned}
\right.
\]
Here $\bar{p}$ is the normalized pressure, and $k,~c$ with subscripts are the normalized coefficients. $\bm W^{(i)}$ and $a^{(i)}$ are the weighting matrix and variables of the $i^{th}$ layer, respectively.

In addition, $\gamma = 1.4$ in \eqref{eq:ideal} is acceptable for most cases, while it is inoperative when the gas undergoes high pressure and temperature. In this paper, we adopt the tabulated form of real gas EOS in \cite{Symbalisty1995} and obtain a similar neural network EOS to \eqref{eq:dnn}. For more information about the values of the above coefficients, please refer to our previous work \cite{Li2023neural}.

% \subsubsection*{\bf Rigid material model}
$\bullet \textbf{ Rigid material model}$

In the phase of solid, the hydrostatic pressure resists the volumetric deformation, while the deviatoric stress tensor resists the shear deformation and obeys Hooke's law in the elastic region and the specific plastic flow rule in the plastic region. In this paper, we ignore the deformation of the solid and treat it as a rigid body model which obeys 
\[
\bm\sigma({\bm\varepsilon})=\bm 0, \quad  \bm \varepsilon = \bm 0.
\]
This assumption is very effective when we only focus on the propagation of blast waves, and the numerical results are usually acceptable compared to the rigorous elastoplastic calculation.

%%% Local Variables: 
%%% mode: latex
%%% TeX-master: "article"
%%% End: 

\section{Numerical schemes}
\label{sec:scheme}
In this section, we will briefly introduce the numerical framework of coupling multiphase flow with simplified elastoplastic solid models, and we recommend that interested readers refer to previous work for more details \cite{Fu2022}. The elastoplastic solid is simplified as a rigid solid in the framework of the MPM. The underlying theory is a multi-material description that has the flexibility to incorporate different numerical descriptions for solid and fluid fields within general solution procedures. By choosing the background grid of the MPM to be the same one as the multiphase Eulerian description, the interactions among all materials can be calculated in a general framework. The multiphase flow is solved under the Eulerian perspective by the FVM and relaxation procedure with an infinite rate. Furthermore, we also apply our recent reconstruction scheme, named the deepMTBVD \cite{huang2024struct}, which shows high resolution, high performance, and high efficiency and is introduced in the following section.
% In this section, we briefly introduce the numerical schemes and procedures of the whole computation system, 
% including the coupling scheme between the multiphase flow and the solid with rigid constitutive models,
% solving the multiphase model in \eqref{eq:cons} by first-order Strang splitting method \cite{strang1968splitting}. 
% Furthremore, based on our previous work, the deepMTBVD scheme \cite{huang2024struct}, which is a low-disspation and high-resolution reconstruction scheme, 
% is also provided in the following subsection. 

\begin{figure}[htbp]
\centering
\subfigure[initial value of solid and fluid]
{\includegraphics[width=0.48\textwidth]{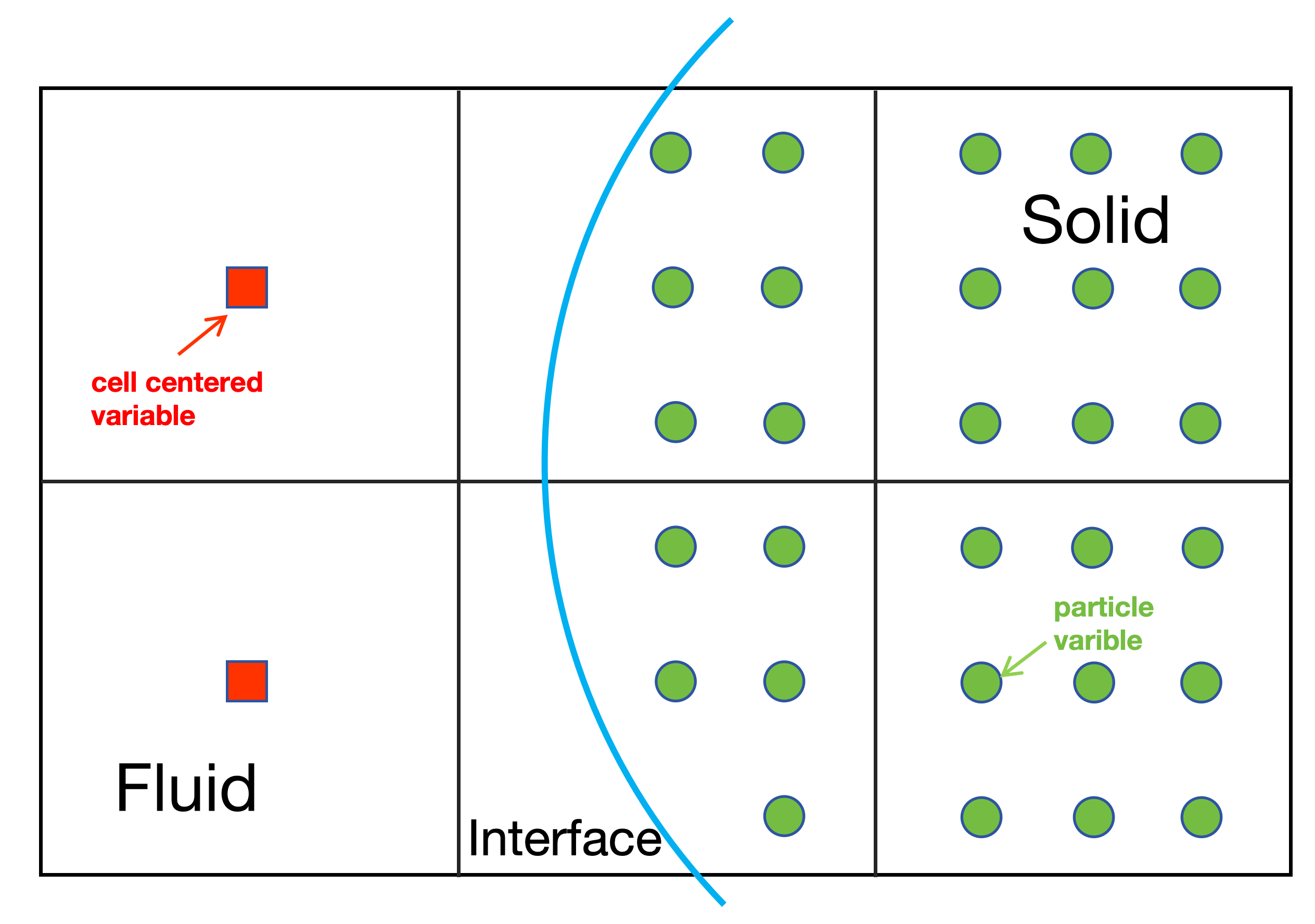}\label{fig:mpmice_cycle:a}}
\quad 
\subfigure[interpolation from particles to grid nodes]
{\includegraphics[width=0.48\textwidth] {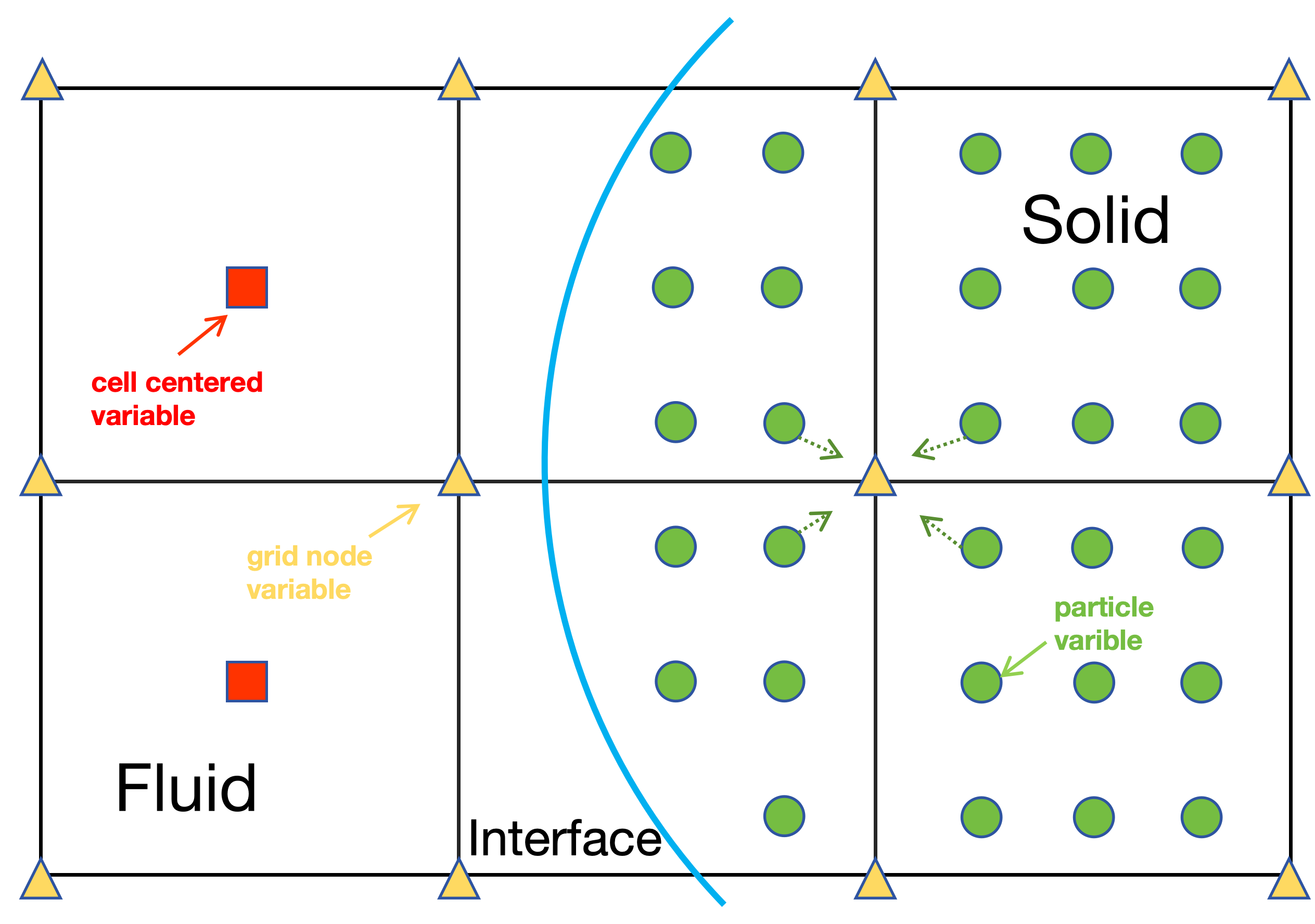}\label{fig:mpmice_cycle:b}}  \\
\vspace{5mm}
\subfigure[interpolation from grid nodes to cell center]
{\includegraphics[width=0.48\textwidth]{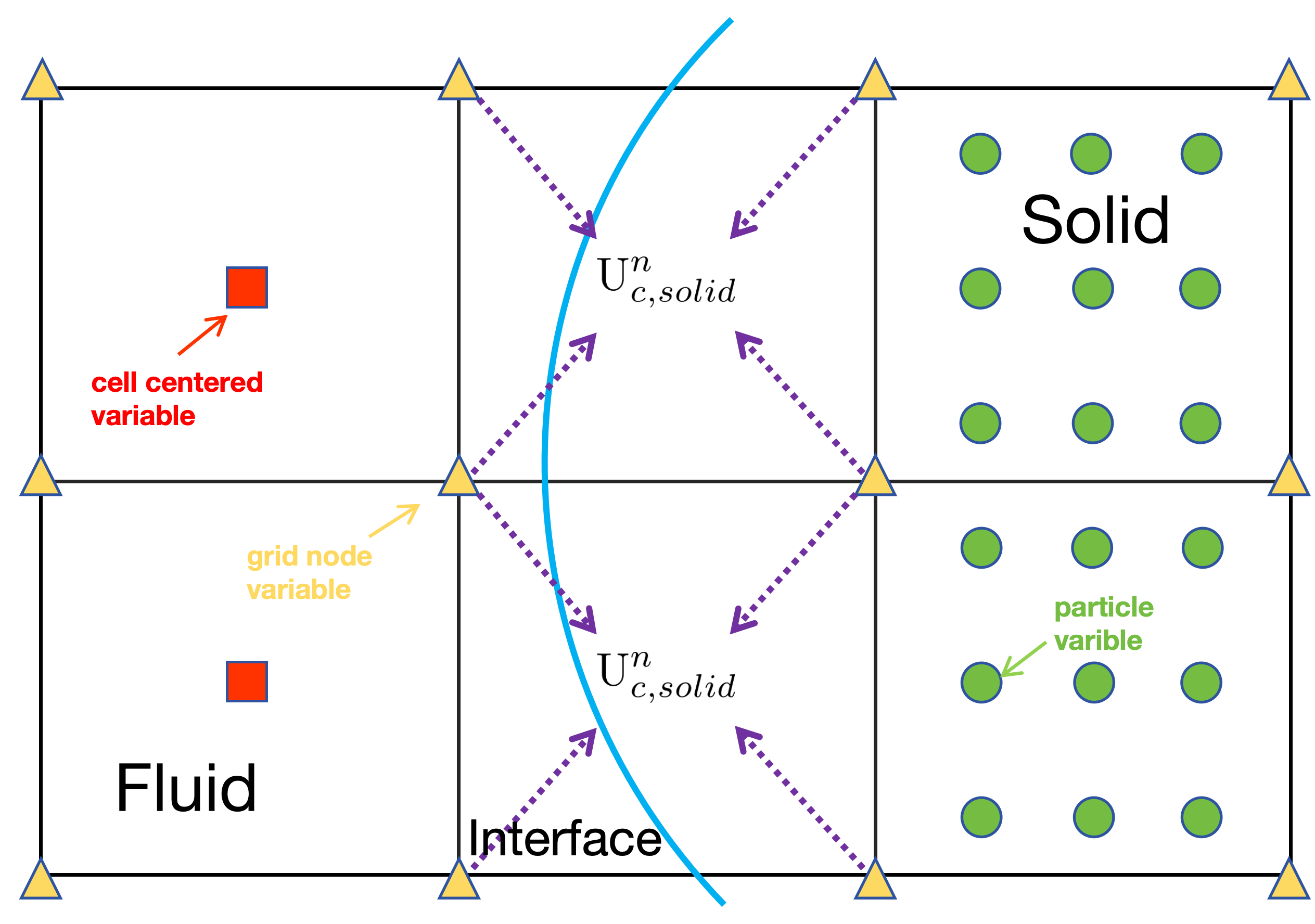}\label{fig:mpmice_cycle:c}} 
\quad
\subfigure[instantaneous velocity and pressure relaxation]
{\includegraphics[width=0.48\textwidth]{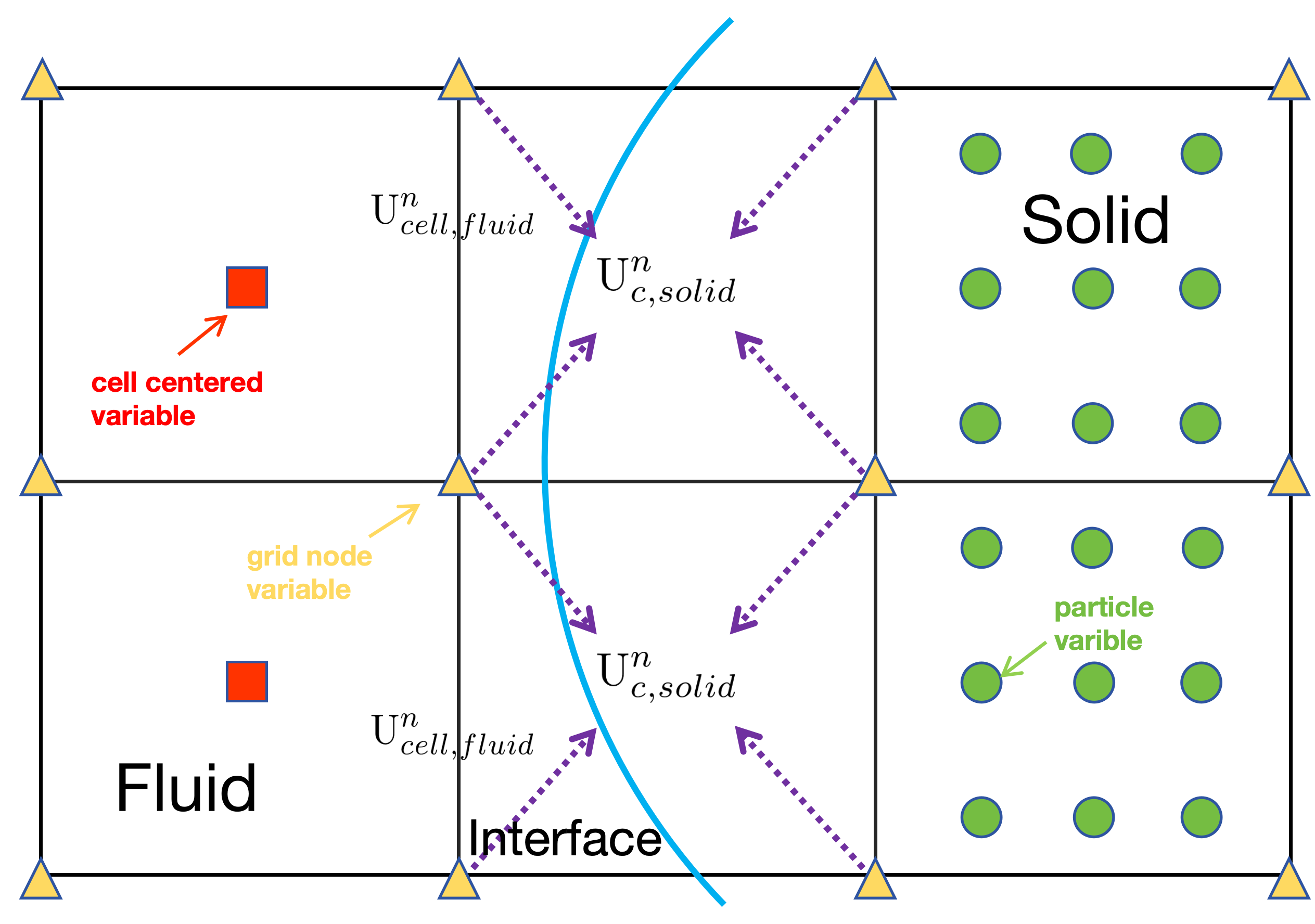}\label{fig:mpmice_cycle:d}}   \\
\vspace{5mm}
\subfigure[interpolation back from cell center to grid nodes]
{\includegraphics[width=0.48\textwidth]{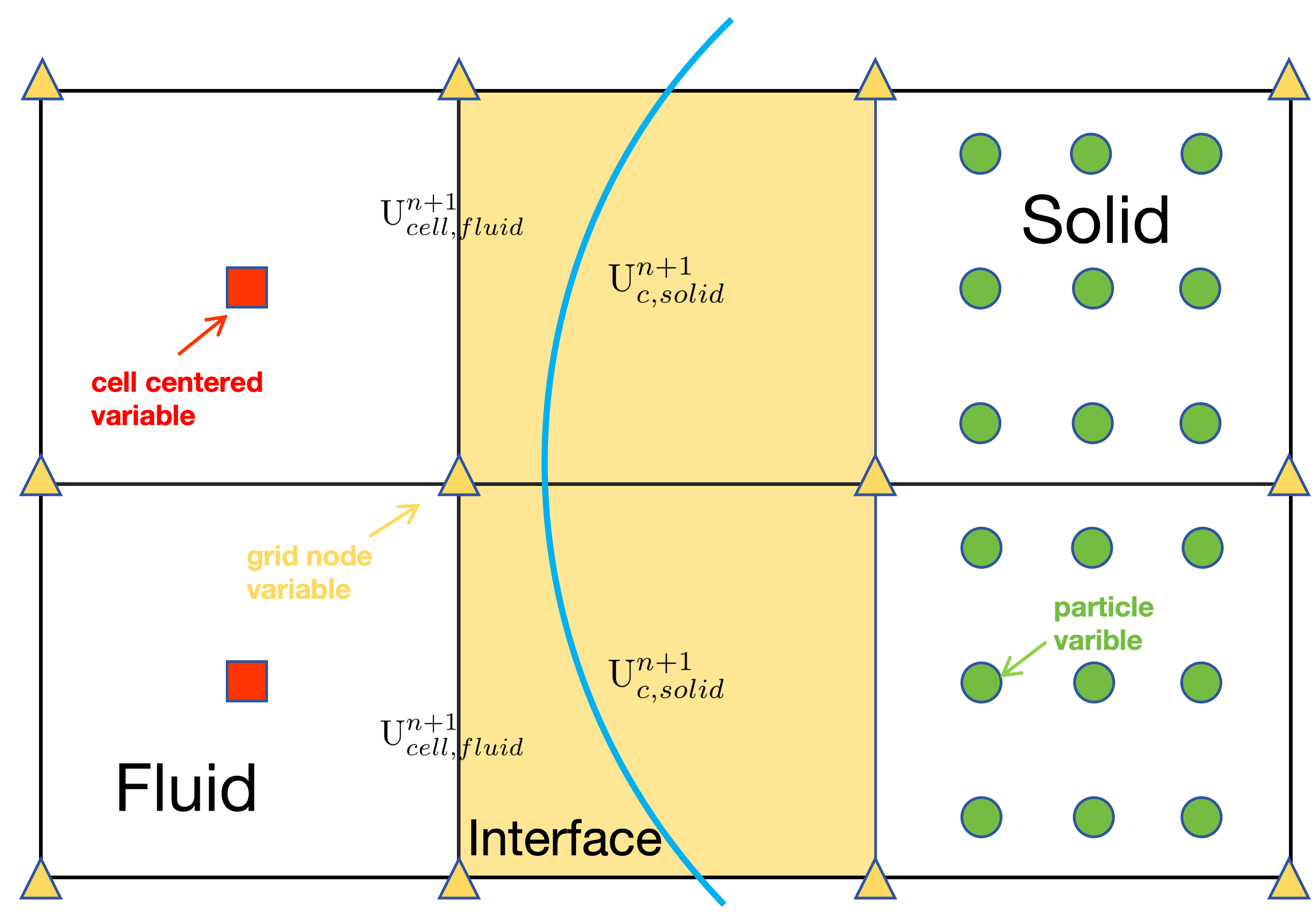}\label{fig:mpmice_cycle:e}} 
\quad
\subfigure[interpolation from grid nodes to nearby particles]
{\includegraphics[width=0.48\textwidth]{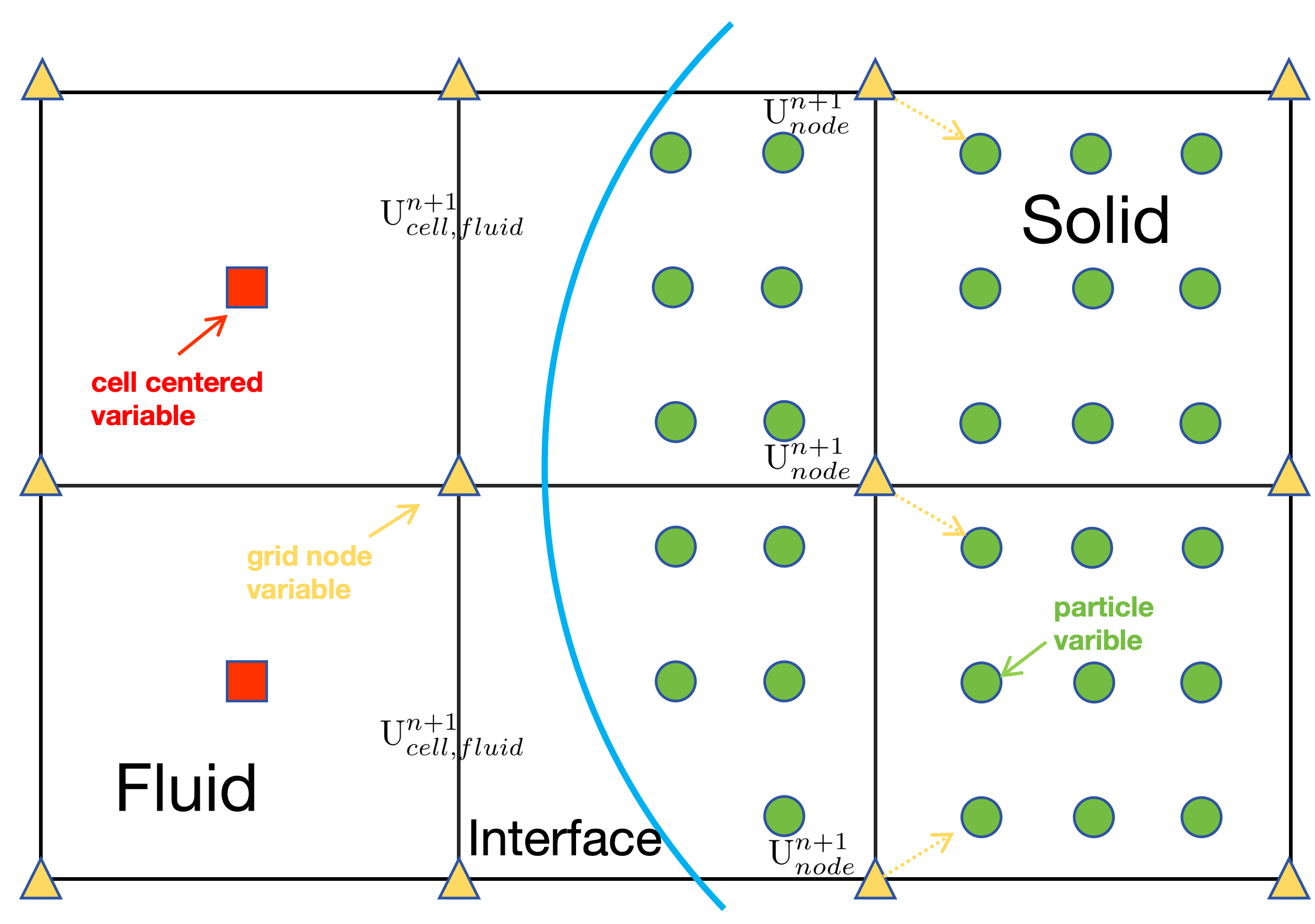}\label{fig:mpmice_cycle:f}}    
\caption{Numerical procedures of the coupling between the multiphase flow and material point method.}
\label{fig:mpmice_cycle}
\end{figure}

At the beginning of the calculation, the phase interface between the fluid and solid is aligned with the boundary of the solid particles, shown in Fig.~\ref{fig:mpmice_cycle:a}, then the initial volume fraction of the fluid in the interfacial cell is computed by using the geometric analysis. After setting the initial values of each material, the homogeneous system of the multiphase flow governing equations \eqref{eq:cons}  is solved by a Godunov-type scheme. A sequence of ordinary differential equations involving the source term of the governing equations is solved by a stiff relaxation technique. The main schemes are listed as follows.

\subsection{Projection of the particle states to the grid nodes} 
The simulation begins by interpolating the particle variables of the solid particles to the grid nodes. As shown in Fig. \ref{fig:mpmice_cycle:b}, the particle variables of the solid are projected to the grid nodes, which follows:
\[
\bm U_{I}^{n} = \sum_{v=1}^{n_{v}} \mathcal{N}_{Iv}^{n} \bm U_{v}^{n} ,
\]
where the subscript $I$ and $v$ represent the index of grid nodes and particles respectively, $n_{v}$ stands for the number of particles, $\mathcal{N}_{Iv}$ stands for the shape function of particle $v$ in the MPM.

In Fig.~\ref{fig:mpmice_cycle:c}, a subsequent projection from the grid nodes to the cell centers is implemented to construct the cell-centered variables of solid and keep the framework of multi-component numerical schemes, which is given as:
\[
\bm{U}_c^n = \sum_{I=1}^{n_I} {\mathcal N}_{ij}^n \bm U_I^n.
\]
Here a linear shape function ${\mathcal N}_{ij}$ is adopted to interpolate the variables between the grid nodes and cell center, $n_I$ is the number of grid nodes in each cell, where $n_I=4$, $\mathcal N_{ij}=1/4$ for two dimensional structured grid, and $n_I=8$, $\mathcal N_{ij}=1/8$ for three-dimensional grid. For the grid nodes at the reflective face or edge, the value of $\mathcal N_{ij}$ is doubled to maintain the consistency of the boundary conditions.

\subsection{Solving the multiphase fluid in system \eqref{eq:cons}}
The elastoplastic solid is considered as one component of the multiphase flow after being interpolated to the cell center. In the Cartesian grid, we consider the discrete method of Eq. $\eqref{eq:cons}$ in $x$-dimension only, $y$-dimension and {z}-dimension are similar. The computational domain is divided into $N$ cells, and denotes $i^{th} $cell $I_{i} = \left[x_{i-1/2}, x_{i+1/2}\right], i \in \{1, \cdots, N\}$. 
Assume that grid size is $\Delta x_{i} = x_{i+1/2}-x_{i-1/2}$ and cell center locates at $x_{i+1/2}$, the general integral formulation of the physical variables $\bm{U}$ over a finite volume cell $I_{i}$, which follows the cell-average values, is given by:
\begin{equation}
    \Bar{\bm{U}}_{i} = \frac{1}{\Delta x_{i}} \int_{x_{i-1/2}}^{x_{i+1/2}}\bm{U} dx.
\end{equation}

When a first-order time integration is employed, we need three sub-steps to reach the desired solution, provided by the Godunov splitting method \cite{godunov1956,strang1968splitting}. Successive integrator gives the solution of the state vector in the next time step:
\begin{equation}
    \Bar{\bm{U}}^{n+1} = \mathcal{L}_{\bm{p}}^{\Delta t} \mathcal{L}_{\bm{u}}^{\Delta t} \mathcal{L}_{\bm{B}}^{\Delta t} \Bar{\bm{U}}^{n}.
\end{equation}
Here $\Delta t$ is the discretizing time step, $\mathcal{L}_{\bm{B}}^{\Delta t}$ is the operator solving the homogeneous system in Eq. \eqref{eq:cons}, 
$\mathcal{L}_{\bm{u}}^{\Delta t}$ and $\mathcal{L}_{\bm{p}}^{\Delta t}$ refers to velocity and pressure relaxation procedures in the source term $\bm{H}(\bm{U})$, respectively.

\subsubsection{Calculation of the homogeneous integrator $\mathcal{L}_{\bm{B}}^{\Delta t}$}
To solve the multiphase model in Eq.~\eqref{eq:cons}, we apply the Godunov-type method in \cite{Saurel2009simple, Saurel2018Diffuse} for the spatial discretization of cell $I_{i}$.
\begin{equation}
    \begin{aligned}
        &\frac{\Bar{U}^{n+\Delta t} - \Bar{U}^{n}}{\Delta t} = \mathcal{L}_{\bm{B}}^{\Delta t}, \\
        & \mathcal{L}_{\bm{B}}^{\Delta t} = -\frac{1}{\Delta x_{i}}\left(\left(\bm{F}_{i+1/2} - \bm{F}_{i-1/2}\right) + \bm{B}(\Bar{\bm{U}}_{i})(\alpha_{i+1/2}^{*} - \alpha_{i-1/2}^{*})\right) .
    \end{aligned}
    % \frac{d}{dt}\Bar{\bm{U}}_{i} = -\left(\frac{1}{\Delta x_{i}}\left(\bm{F}_{i+1/2} - \bm{F}_{i-1/2}\right) + \bm{B}(\Bar{\bm{U}}_{i})(\alpha_{i+1/2}^{*} - \alpha_{i-1/2}^{*})\right) + \bm{H}(\Bar{\bm{U}}_{i}).
    \label{discrete}
\end{equation}
where $\bm{F}_{i-1/2}, \bm{F}_{i+1/2}$ denote the right-side and left-side flux on the boundary on cell $I_{i}$, which are described in detail:
\begin{equation}
    \bm{F}_{i-1/2} = \bm{F}^{*}(\bm{U}_{i-1/2}^{L}, \bm{U}_{i-1/2}^{R}), \quad \bm{F}_{i+1/2} = \bm{F}^{*}(\bm{U}_{i+1/2}^{L}, \bm{U}_{i+1/2}^{R}). \label{riemann:a} 
\end{equation} 
The subscript $``*"$ in \eqref{discrete} and \eqref{riemann:a} refers to the Riemann solution and flux calculated from the left and right state between the cell interfaces. Both sides states $\bm{U}^{L}$ and $\bm{U}^{R}$ are obtained from the specific reconstruction schemes at the cell interfaces, which will be discussed in the following subsection. In this work, we solve the Riemann problem by the HLLC Riemann solver, whose specific formulation can be found in \cite{torobook, Saurel2009simple}.
% \begin{equation}
%     \frac{d}{dt} \Bar{\bm{U}}_{i} = -\frac{1}{\Delta x_{i}}(\mathcal{A}^{+}\Delta\bm{U}_{i-1/2} + \mathcal{A}^{-}\Delta\bm{U}_{i+1/2} + \mathcal{A}\Delta\bm{U}_{i}),
%     \label{numerical:fvm}
% \end{equation}
% where $\mathcal{A}^{+}\bm{U}_{i-1/2}$ and $\mathcal{A}^{-}\bm{U}_{i+1/2}$ are right-moving and left-moving fluctuations and $\mathcal{A}\bm{U}_{i}$ is the total fluctuation of cell $I_{i}$. In detail, right-moving and left-moving fluctuations can be determined by corresponding Riemann problems:
% \begin{equation}
%     \mathcal{A}^{\pm}\Delta\bm{U}_{i\mp1/2} = \sum_{k=1}^{K}\left[s^{k}(\bm{U}_{i-1/2}^{L}, \bm{U}_{i+1/2}^{R})\right]^{\pm} \bm{W}^{k}(\bm{U}_{i-1/2}^{L}, \bm{U}_{i+1/2}^{R}),
%     \label{numerical:wave_propagation}
% \end{equation}
% where $K$, $s^{k}$ and $\bm{W}^{k}$ are the number of waves, moving speeds, and jumps of three propagation discontinuities solved by Riemann solvers such as HLLC Riemann solver \cite{torobook}, respectively. $\bm{U}_{i-1/2}^{L}$ and $\bm{U}_{i-1/2}^{R}$ are reconstruction values from neighboring cells by specific schemes at the interface $x = x_{i-1/2}$. The reconstruction method is the core of this paper, which will be discussed in the next section. Similarly, the total fluctuation can be calculated by 
% \begin{equation}
%     \mathcal{A}\Delta\bm{U}_{i} = \sum_{k=1}^{K}\left[s^{k}(\bm{U}_{i-1/2}^{R}, \bm{U}_{i+1/2}^{L})\right]^{\pm} \bm{W}^{k}(\bm{U}_{i-1/2}^{R}, \bm{U}_{i+1/2}^{L}),
%     \label{numerical:total_fluctuation}
% \end{equation}

Once the spatial discretization is given, we employ the forward Euler scheme for three-dimensional multiphysics numerical tests:
\begin{equation}
    \Bar{\bm{U}}^{n+1} = \Bar{\bm{U}}^{n} + \Delta t \mathcal{L}_{\bm{B}}^{\Delta t}(\Bar{\bm{U}}^{n}).
    \label{numerical:timeintegral}
\end{equation}
Furthermore, we also apply the second-order strong stability-preserving (SSP) \cite{gottlieb2001ssp} scheme for the single phase in numerical simulation:
\begin{equation}
    \begin{aligned}
    & \bm{U}^{*} = \Bar{\bm{U}}^{n} + \Delta t \mathcal{L}_{\bm{B}}^{\Delta t}(\Bar{\bm{U}}^{n}),\\
    & \Bar{\bm{U}}^{n+1} = \frac{1}{2}\Bar{\bm{U}}^{n} + \frac{1}{2}(\bm{U}^{*} + \Delta t \mathcal{L}_{\bm{B}}^{\Delta t}(\bm{U}^{*})).
    \end{aligned}
\end{equation}

% \subsubsection{Instantaneous relaxtion of pressure and velocity}
\subsubsection{Calculation of the relaxation operator of velocity $\mathcal{L}_{\bm{u}}^{\Delta t}$ and pressure $\mathcal{L}_{\bm{p}}^{\Delta t}$}
After solving the homogeneous system, we solve a sequence of ordinary differential equations accounting for the relaxation source terms of system \eqref{eq:cons}. Since the non-equilibrium effects of pressure and velocity are not the focus of our attention from a physical perspective, we assume that the parameters $\mu$ and $\lambda$ are infinite, which means that the two phases reach equilibrium in an instant. For the instantaneous relaxation of pressure and velocity, we solve the following system of ordinary differential equations:
\[
\dfrac{\partial \bm{U}}{\partial t} = \bm H_{\bm{u}}(\bm U) + \bm H_{\bm{p}}(\bm U), 
\]
with $\mu \to \infty$ and $\lambda \to \infty$, where
$\bm H_{\bm{u}}(\bm U)=[0,~0,~0,~\lambda (\bm u_2-\bm u_1),~-\lambda (\bm u_2-\bm u_1),~\lambda \bm u_I(\bm u_2-\bm u_1),~-\lambda \bm u_I(\bm u_2-\bm u_1) ]^{\top}$ is the velocity relaxation term, and $\bm H_{\bm{p}}(\bm U)=[\mu(p_1-p_2),~0,~0,~0,~0,~\mu p_I(p_2-p_1),~-\mu p_I(p_2-p_1) ]^{\top}$  is the pressure relaxation term.

% \subsubsection*{Instantaneous velocity relaxation}
$\bullet \textbf{ Instantaneous velocity relaxation operator} \mathcal{L}_{\bm{u}}^{\Delta t}$

For the instantaneous velocity relaxation ($\lambda \rightarrow \infty$), the ordinary differential equation $\partial \bm U/\partial t=\bm H_{\bm{u}}(\bm U)$ leads to 
\[
\begin{aligned}
\dfrac{\partial \bm u_1}{\partial t} &=  \dfrac{\lambda(\bm u_1-\bm u_2)}{\alpha_1 \rho_1}, \\
\dfrac{\partial \bm u_2}{\partial t} &= -\dfrac{\lambda(\bm u_1-\bm u_2)}{\alpha_2 \rho_2}.
\end{aligned}
\]
After some simple manipulation, the volume fraction $\alpha_{i}$ and the density $p_{i}$ of each phase is obtained by taking the method in \cite{Saurel1999multiphase}
\begin{equation*}
    \alpha_{i} = \alpha_{i}^{0}, \quad p_{i} = p_{i}^{0}, \quad i = 1, 2.
\end{equation*}
The relaxed solution of the velocities satisfies 
\[
\bm{u}_{I}=\bm{u}_{1}=\bm{u}_{2}=
  \dfrac{\alpha_1 \rho_1 \bm u_{1}^{0} + \alpha_2 \rho_2 \bm u_{2}^{0}} {\alpha_1 \rho_1 + \alpha_2 \rho_2},
\]
and the relaxed internal energy of each phase has the following value 
\[
e_i=e_{i}^{0} \pm \dfrac{1}{2} (\bm u_i-\bm u_{i}^{0})\cdot(\bm u_i-\bm u_{i}^{0}), \quad i=1,2.
\]
Here, $\alpha_{i}^{0}, p_{i}^{0}$, $e_{i}^{0}$ and $\bm u_{i}^{0}$ are respectively the volume fraction, pressure,
internal energy and velocity for $i^{th}$ phase before relaxation, which perform as the initial value 
of the ODE system.

% \subsubsection*{Instantaneous pressure relaxation operator $\mathcal{L}_{\bm{u}}^{\Delta t}$}
$\bullet \textbf{ Instantaneous pressure relaxation operator} \mathcal{L}_{\bm{p}}^{\Delta t}$

For the instantaneous pressure relaxation ($p \rightarrow \infty$), the ordinary differential equation $\partial \bm U/\partial t=\bm H_{\bm{p}}(\bm U)$ leads to 
\[
\dfrac{\partial e_i}{\partial t} + p_I \dfrac{\partial v_i}{\partial t} = 0, \quad  i=1,2,
\]
where $v_i=1/\rho_i$ stands for the specific volume of phase $i$. In order to get the approximate solution, we integrate the above equation, which yields the results:
\begin{equation}
\label{eq:internal_relaxation2}
e_i^{\star}(p^{\star},v_i^{\star}) - e_i(p^0, v_i^0) + \bar{p}_I (v_i^{\star} - v_i^0) = 0,
\end{equation}
where the superscript $``0",~``\star"$ represent the value before and after pressure relaxation respectively. $\Bar{p}_{I}$ represents the mean interfacial pressure between the initial state $``0"$ and relaxed state $``\star"$. The possible estimates of $\bar{p}_{I}$ are $p_{I}^{0}$ or $p_{I}^{\star}$, which denote the initial and relaxed pressure respectively.
According to the results in literature \cite{Saurel2009simple}, 
the computation results between either estimate show negligible difference.
Here we choose $\bar{p}_I=(p_I^{0}+p_I)/2$, and $e_i^{\star},~e_i$ are calculated by using the corresponding equations of state for each phase.

We solve the following algebraic equation with a single pressure $p$, by using the Newton-Raphson method
\[
\alpha^0_1 \rho_1^0 v_1(p) + \alpha^0_2 \rho_2^0 v_2(p) = 1,
\]
where $\alpha^0_1, \alpha^0_2$ and $\rho_1^0, \rho_2^0$ are the phasic volume fractions and densities before relaxation, and $v_1=1/\rho_1, v_2=1/\rho_2$ are the specific volume after relaxation.
Here, $\alpha_{i}\rho_{i}$ remains constant during the pressure relaxation. Thus, the system can be reformulated 
by a single equation with a single unknown variable ($p$), which is ultimately solved by the iterative method within several steps. For more details, please refer to \cite{Zein2010modeling, Saurel2009simple}.
Once the relaxation pressure $p^{\star}$ is determined, the internal energy of the each phase $e_{i}^{\star}$ is obtained by the corresponding
EOS before advancing to the next time step:
\[
e_{i}^{\star} = e_{i}(\alpha_{i}, \alpha_{i}\rho_{i}, p^{\star}).
\]

\subsection{Advancement of particle position and variables}
After advancing the hyperbolic operator and finishing the relaxation procedures, the changes in mass, momentum and total energy of solid 
computed at cell centers are interpolated back to the neighboring nodes and particles progressively, as shown in 
Fig. \ref{fig:mpmice_cycle:e} and Fig. \ref{fig:mpmice_cycle:f}, respectively. For details in these steps, we recommend that interested readers refer to \cite{Fu2022}. Due to the assumption of the rigid body 
of solid, the position and values of each variable for solid particles remain constant and play the role of wall boundary conditions.
\begin{remark}
When all the procedures above for the one-time step are finished, we put the simulation into the next step.
\end{remark}

\subsection{The deepMTBVD reconstruction scheme}\label{subsec:deepmtbvd}
%The solutions with first-order accuracy of advection will smear out the contact discontinuities. In this section, we adopt the novel low-dissipation reconstruction scheme based on the artificial neural network, the deepMTBVD scheme \cite{huang2024struct}, to the compressible multi-phase flows system. In this subsection, we briefly introduce the deepMTBVD scheme in one dimension, which can be directly extended to multiple dimensions using dimension-splitting techniques. We first present the boundary variation diminishing (BVD) principle and then introduce the deepMTBVD scheme later.
% For simplicity, we briefly introduce the deepMTBVD scheme in one dimension. Our work focuses on Cartesian grids, which can be directly extended to multiple dimensions using dimension-splitting techniques. The computational domain is divided into $N$ cells, and denotes $i^{h}$ cell $I_{i} = [x_{i−1/2}, x_{i+1/2}], i \in \{1, \cdots, N\}$. Assume that grid size is $\Delta x_{i} = x_{i+1/2} − x_{i−1/2}$,  and cell center locates at $x_{i+1/2}$, the general integral formulation of the conservative variable U over a finite volume cell $I_{i}$, which follows the cell-average values, is given by:
% \begin{equation*}
%     \Bar{\bm{U}}_{i} = \frac{1}{\Delta x_{i}} \int_{x_{i-1/2}}^{x_{i+1/2}}\bm{U} dx.
% \end{equation*}

In this subsection, we briefly introduce the high-fidelity deepMTBVD scheme\cite{huang2024struct} used to reconstruct the left and right stcomplexityate at each cell interface for calculating numerical fluxes in Eq. \eqref{riemann:a}. It is a novel low-dissipation reconstruction scheme for compressible single- and multi-phase flows based on the MUSCL-THINC-BVD scheme and an artificial neural network.

For the sake of simplicity, we drop the subscript signal in \eqref{riemann:a}. Assume the left and right sides of an arbitrary cell boundary are $\bm{U}^{L}$ and $\bm{U}^{R}$, the approximate Riemann solver can be reformulated as canonical formulation as follows \cite{sun2016bvd}:
\begin{equation}
    \bm{F}(\bm{U}^{L}, \bm{U}^{R}) = \frac{1}{2}\left(\bm{F}(\bm{U}^{L}) + \bm{F}(\bm{U}^{R})\right) - \bm{A}(\bm{U}^{L}, \bm{U}^{R})(\bm{U}^{R}-\bm{U}^{L}). \label{numerical:method:canonical}
\end{equation}

The right-hand side of Eq. \eqref{numerical:method:canonical} is divided into central and dissipative parts. Here, $\bm{A}(\bm{U}^{L}, \bm{U}^{R})$ is a system matrix computed from $\bm{U}^{L}$ and $\bm{U}^{R}$, and $\left|\bm{U}^{R}-\bm{U}^{L}\right|$ is referred to as the boundary variation. This dissipative part intrinsically introduces excessive diffusion in the numerical solution, which leads to the BVD principle \cite{sun2016bvd}. The fundamental idea of the BVD principle is that minimizing boundary variation can reduce numerical dissipation in flux calculations \cite{sun2016bvd}. As described in $\eqref{numerical:method:candidate}$, a series of BVD algorithm \cite{cheng2021low, deng2018high, DengJiang-848, DengShimizu-142, HouZhao-1564, wakimura2022sym, DengXie-99} are designed to select the optimal reconstruction function $Q_{i}^{\xi_{opt}}$ from the candidates set $\Xi_{i}$ to minimize the dissipative term on cell $I_{i}$.

\begin{equation}
    \Xi_{i} = \left\{Q_{i}^{\xi_{1}}, Q_{i}^{\xi_{2}}, \cdots, Q_{i}^{\xi_{n}}\right\} \xrightarrow{\text{BVD algorithm}} Q_{i}^{\xi_{opt}}. \label{numerical:method:candidate}
\end{equation}

BVD schemes utilize polynomial-based methods in smooth regions and the THINC scheme to capture discontinuities, effectively reducing overall numerical dissipation. Traditional BVD schemes require preparing all candidate interpolation functions before selecting the optimal one for each cell. In contrast, the deepMTBVD scheme, leveraging an artificial neural network, pre-selects the most suitable interpolation function prior to reconstruction. This approach allows for a single reconstruction per cell, enhancing efficiency and demonstrating greater flexibility and superiority compared to conventional BVD schemes.
\begin{figure}[htbp]
    \centering
    \includegraphics[width=0.5\linewidth]{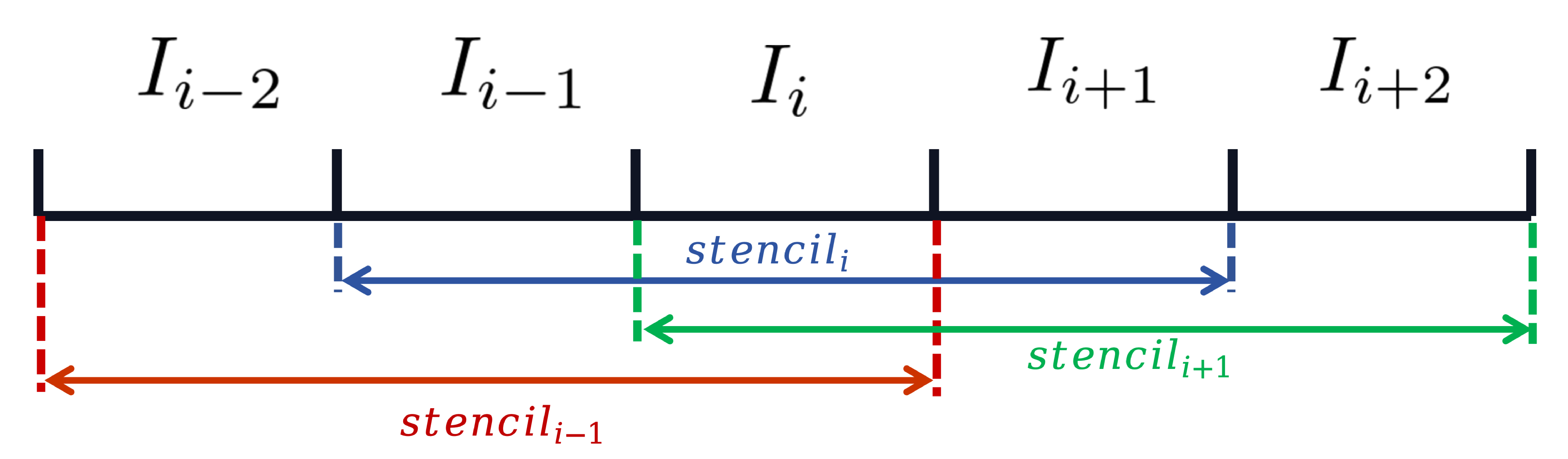}
    \caption{The stencils of the deepMTBVD schemes, duplicated from \cite{huang2024struct}.}
    \label{fig:numerical:method:stencil}
\end{figure}

The candidate set $\Xi_{i}$ of the deepMTBVD scheme involves the MUSCL and THINC schemes. As shown in Fig. \ref{fig:numerical:method:stencil}, the stencil of the MUSCL and the THINC schemes consists of neighboring cells of $I_{i}$. Thus, the stencil of the BVD algorithm covers three different stencils, including $``stencil_{i-1}", ``stencil_{i}", \text{ and } ``stencil_{i+1}"$, and is denoted as $S_{i} = \left\{\Bar{u}_{i-2}, \Bar{u}_{i-1}, \Bar{u}_{i}, \Bar{u}_{i+1}, \Bar{u}_{i+2}\right\}$. According to the lemma in \cite{huang2024struct}, we reformulate the stencil data $S_{i}$ as follows:
\begin{equation}
    \Tilde{u}_{i} = \left\{
    \begin{aligned}
    \frac{\Bar{u}_{i} - \Tilde{u}_{\min}}{\Tilde{u}_{\max} - \Tilde{u}_{\min}}, & \quad |\Tilde{u}_{\max} - \Tilde{u}_{\min}| \ge \zeta \text{ and } \chi_{i} = 1,\\
    0, & \quad |\Tilde{u}_{\max} - \Tilde{u}_{\min}| < \zeta \text{ or } \chi_{i} = 0.
    \end{aligned}
    \label{numerical:method:reformulate}
    \right.
\end{equation}
Here $\Tilde{u}_{\max} = \max\{\bar{u}_{i-2}, \bar{u}_{i-1}, \bar{u}_{i}, \bar{u}_{i+1}, \bar{u}_{i+2}\} $ and $\Tilde{u}_{\min} = \min\{\bar{u}_{i-2}, \bar{u}_{i-1}, \bar{u}_{i}, \bar{u}_{i+1}, \bar{u}_{i+2}\}$, which is the maximum and minimum of stencil S, respectively. $\zeta = 10^{-15}$ is a small positive number. $\chi_{i}$ denotes the monotone indicator, which is defined as follows:
\begin{equation}
    \chi_{i} = \left\{
    \begin{aligned}
        0, & \quad (\bar{u}_{i} - \bar{u}_{i-1})(\bar{u}_{i+1}-\bar{u}_{i}) < 0, \\
        1, & \quad \text{otherwise}.
    \end{aligned}
    \right.
\end{equation}

\begin{figure}[htbp]
    \centering
    \includegraphics[width=0.7\linewidth]{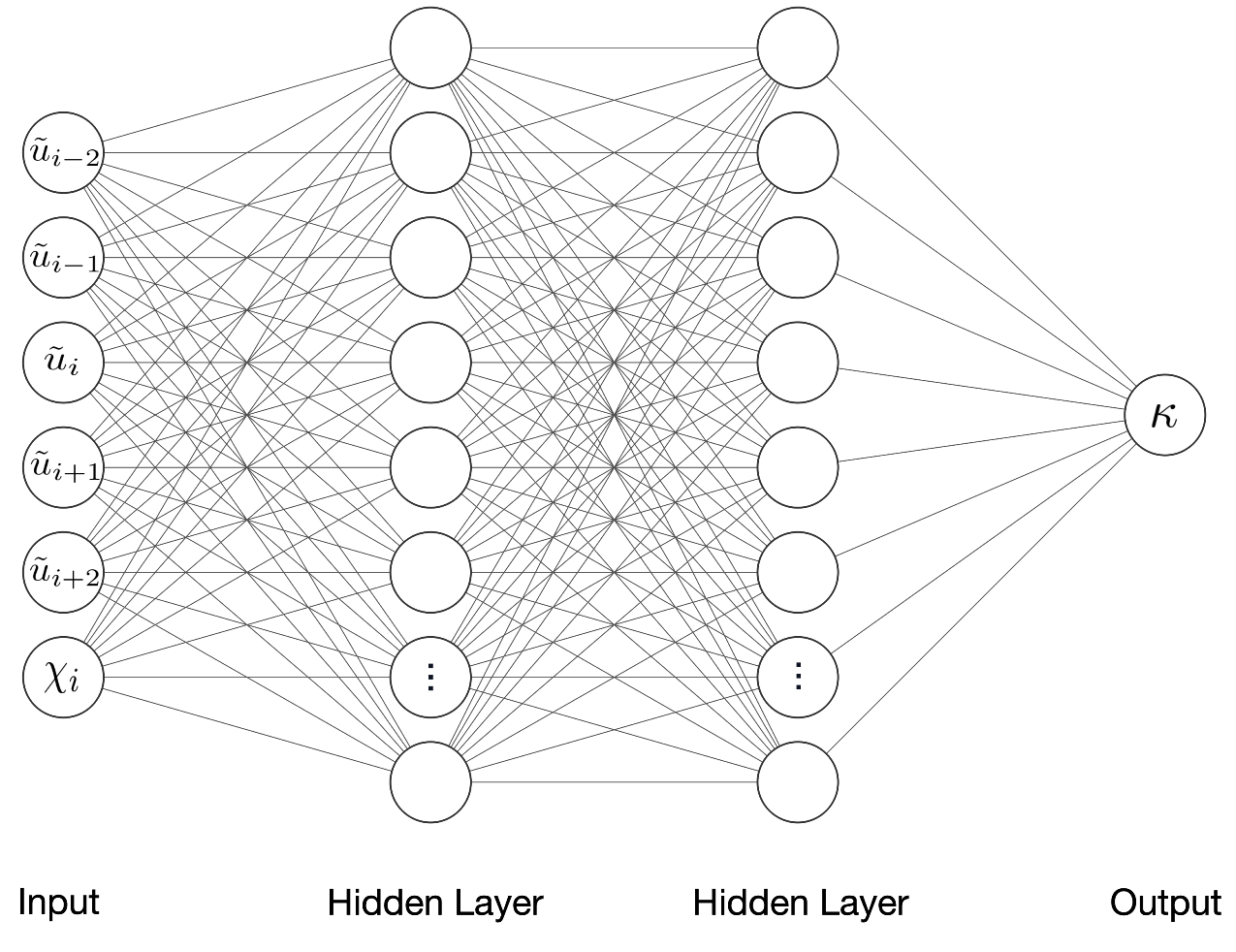}
    \caption{The structure of neural network}
    \label{fig:numerical:method:network}
\end{figure}

We present a multi-layer perceptron architecture as in Fig. \ref{fig:numerical:method:network}, a fully connected neural network with six inputs, several hidden layers, and one output. Following the discussion above, we transform the primary data of $S_{i}$ to $ \Tilde{u}_{i-2}, \Tilde{u}_{i-1}, \Tilde{u}_{i}, \Tilde{u}_{i+1}, \Tilde{u}_{i+2}$ and monotone indicator $\chi_{i}$ of cell $I_{i}$ as preliminary inputs before reformulating by \eqref{numerical:method:reformulate}. The output of the neural network, $\kappa$, is the chosen probability of the THINC scheme. In practice, it is recommended to define the reference $\kappa_{ref}$ artificially to identify the specific reconstruction scheme on a given cell, which follows:
\begin{equation*}
    Q_{i} = \left\{
    \begin{aligned}
        MUSCL, & \quad \kappa < \kappa_{ref}, \\
        THINC, & \quad \text{otherwise}.
    \end{aligned}
    \right.
\end{equation*}

In brief, we show the framework of applying the offline-trained indicator deepMTBVD to the reconstruction step. The implementation of deepMTBVD is detailed in the Algorithm. \ref{numerical:method:deepmtbvd}.

\IncMargin{1em}
\begin{algorithm}[ht] 
\SetKwData{Left}{left}\SetKwData{This}{this}
\SetKwData{Up}{up} \SetKwFunction{Union}{Union}
\SetKwFunction{deepmtbvd}{$\text{deepMTBVD}$} 
\SetKwFunction{secondssp}{$\text{second-order SSP Runge-Kutta}$}
\SetKwInOut{Input}{input}
\SetKwInOut{Output}{output}
    \Input{Physical variable $\Bar{u}_{i}^{n}$ of $I_{i}$ at time $t^{n}$, including density $\rho$, velocity $\bm{V}$,pressure $p$, volume fraction $\alpha_{k}$.} 
    \Output{$\Bar{u}_{i}^{n+1}$ of each $I_{i}$ at time $t^{n+1}$}
    1. Select input stencil $\left\{\Bar{u}^{n}_{i-2}, \Bar{u}^{n}_{i-1}, \Bar{u}^{n}_{i}, \Bar{u}^{n}_{i+1}, \Bar{u}^{n}_{i+2}\right\}$ and then reformulate it to $\left\{\Tilde{u}_{i-2}, \Tilde{u}_{i-1}, \Tilde{u}_{i}, \Tilde{u}_{i+1}, \Tilde{u}_{i+2}, \chi_{i}\right\}$ for each cell $I_{i}$;
    
    2. Calculate output $\kappa_{i}$ by the deepMTBVD and select the reconstruction scheme by comparing it with reference $\kappa_{ref}$;
    
    3. Obtain the reconstruction value on the cell boundary for each $I_{i}$ and calculate flux by Riemann solver;
   
    4. Repeat steps (1)-(3) in each sub-time step. 
    \caption{Application of deepMTBVD method}
    \label{numerical:method:deepmtbvd}  
\end{algorithm}
\DecMargin{1em}

%%% Local Variables: 
%%% mode: latex
%%% TeX-master: "article"
%%% End: 

\section{Numerical Results} \label{sec:results}
In this section, we provide some numerical examples to validate the numerical method. 
These examples not only cover one-dimensional, two-dimensional, and three-dimensional cases, but also include single-phase Euler equations and multiphase flow equations. To further validate the effectiveness of the algorithm, the computational results also include spherical symmetry, axial symmetry, and explosion waves in large-scale complex urban environments.
One-dimensional simulations are conducted on a uniform interval mesh, while two-dimensional and three-dimensional simulations are carried out on meshes composed of rectangular and hexahedral cells, respectively.

\subsection{One-dimensional problems}
In this part, we present some numerical examples of one-dimensional Riemann problems. The ideal gas law is used with the ratio of specific heats of $\gamma = 1.4$. The HLLC Riemann solver \cite{torobook} is used for computing the numerical flux across cell interfaces. The Courant number is set to 0.4, and acceptable $\kappa_{ref}$ is set to 0.45 unless otherwise stated. Numerical results show that the deepMTBVD scheme performs comparably or even better than the MUSCL-THINC-BVD scheme in capturing the shock wave and contact discontinuity. Furthermore, the deepMTBVD scheme achieves higher computational efficiency than the MUSCL-THINC-BVD scheme.

\subsubsection{Sod's Problem}
\begin{figure}[ht!]
    \centering
    \includegraphics[width=\linewidth]{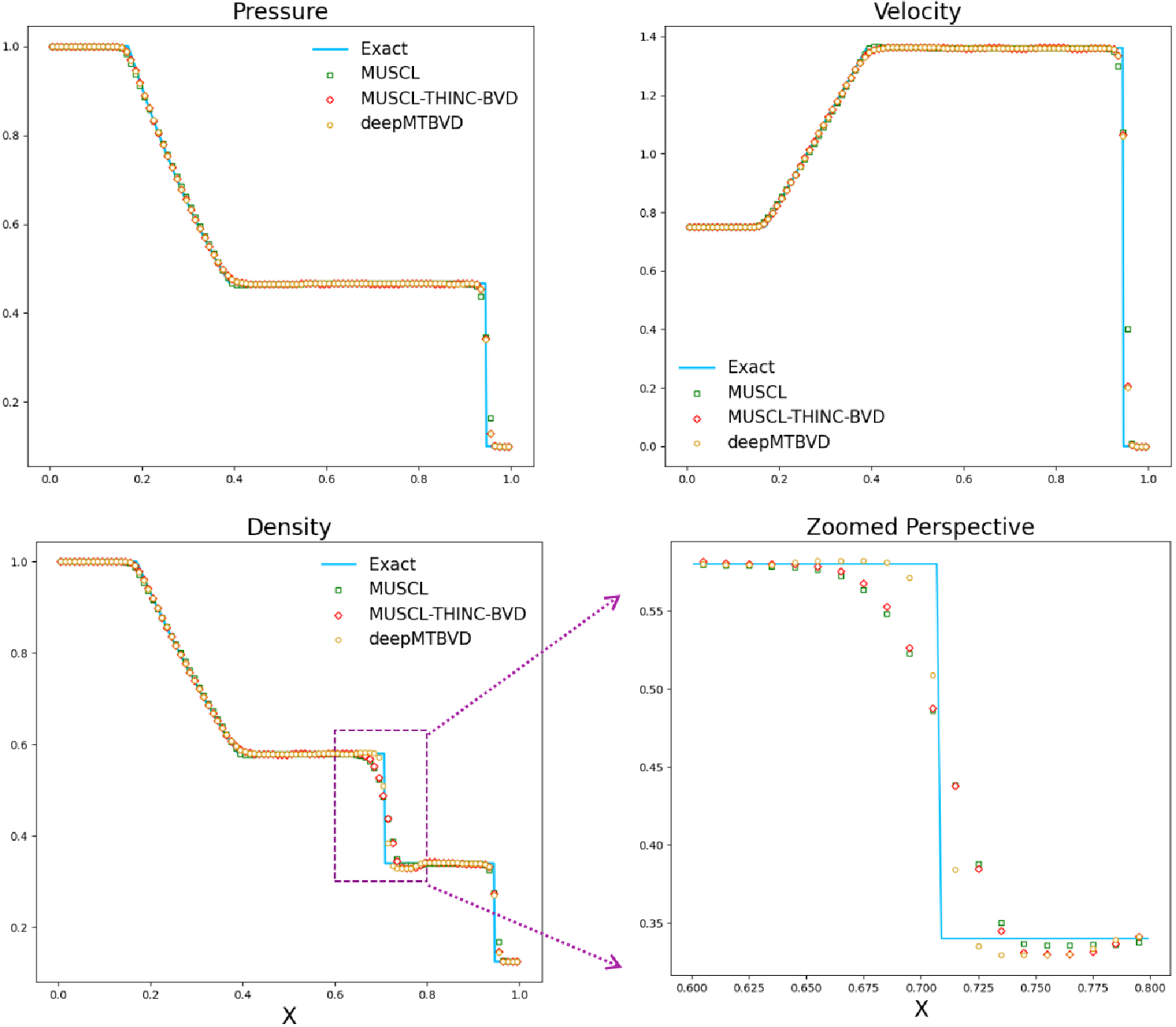}
    \caption{Numerical results of Sod’s problem for velocity, pressure, and density at time $t = 0.3$ with 100 mesh cells. The solid blue
    line is the exact solution, the green square symbol denotes the MUSCL scheme, the red diamond denotes the MUSCL-THINC-BVD scheme and the golden circle denotes deepMTBVD. The right bottom is a zoomed perspective of the dashed square.}
    \label{fig:resutls:case1}
\end{figure}

The Sod problem is widely used to test the shock-capturing scheme and is applied here to evaluate the ability of numerical schemes to resolve shock waves and contact discontinuity. The computational domain is a unit-long tube, and the initial condition is given by \cite{torobook}:
\begin{equation*}
    (\rho, u, p) = \left\{
        \begin{aligned}
            & (1.0, 0.75, 1.0),  & \quad x < 0.3, \\
            & (0.125, 0.0, 1.0), & \quad \text{ otherwise}.
        \end{aligned}\right .
\end{equation*}
We present the numerical results at time $t = 0.3$ with 100 uniform mesh cells. The numerical results from MUSCL, MUSCL-THINC-BVD, and deepMTBVD schemes are presented in Fig. \ref{fig:resutls:case1}. From the zoomed-in view in the lower right of Fig. \ref{fig:resutls:case1}, it is clear that all three schemes resolve the shock wave within two cells similarly. However, the deepMTBVD scheme captures the contact discontinuity with fewer cells compared to MUSCL and MUSCL-THINC-BVD. This indicates that the deepMTBVD scheme exhibits lower numerical dissipation around contact discontinuities, outperforming both the MUSCL and MUSCL-THINC-BVD schemes in this aspect.

\subsubsection{Strong Lax's Problem}
This is another widely used benchmark for testing the performance of the current scheme on the problem with strong discontinuities. 
The initial condition is given by \cite{torobook}:
\begin{equation*}
    (\rho, u, p) = \left\{
        \begin{aligned}
            & (1.0, -19.59745, 1000.0),  & \quad x < 0.8, \\
            & (1.0, -19.59745, 0.01), & \quad \text{otherwise}.
        \end{aligned}\right .
\end{equation*}
%The computational domain is $[0,1]$ with 150 uniform cells. The density of the exact and numerical results at time $t = 0.012$ are shown in Fig. \ref{fig:results:case2}. In Fig. \ref{fig:results:case2:a}, the deepMTBVD resolves the shock wave sharper than the MUSCL scheme. \cld{Furthermore, in the head of the rarefaction wave, the deepMTBVD reproduces the less dissipative solution than the MUSCL scheme.} Compared to the MUSCL-THINC-BVD scheme, the deepMTBVD scheme resolves the contact discontinuity and shock waves similarly. In contrast, the MUSCL-THINC-BVD scheme generates the excessive anti-dissipative solution in the head of the rarefaction. At the same time, the deepMTBVD reproduces the smooth profile, which reveals the robustness and superiority of the deepMTBVD scheme. Furthermore, Fig. \ref{fig:result:laxtime} underlines the time-consuming in this example, from which we find two conclusions. First, the deepMTBVD scheme reduces the reconstruction time, although it increases the selection time since the inference by a neural network. Second, the deepMTBVD scheme spends less time and reproduces a higher-quality numerical solution than the MUSCL-THINC-BVD scheme.

\begin{figure}[ht!]
    \centering
    \subfigure[Comparison of MUSCL and deepMTBVD]{\includegraphics[width=0.49\linewidth]{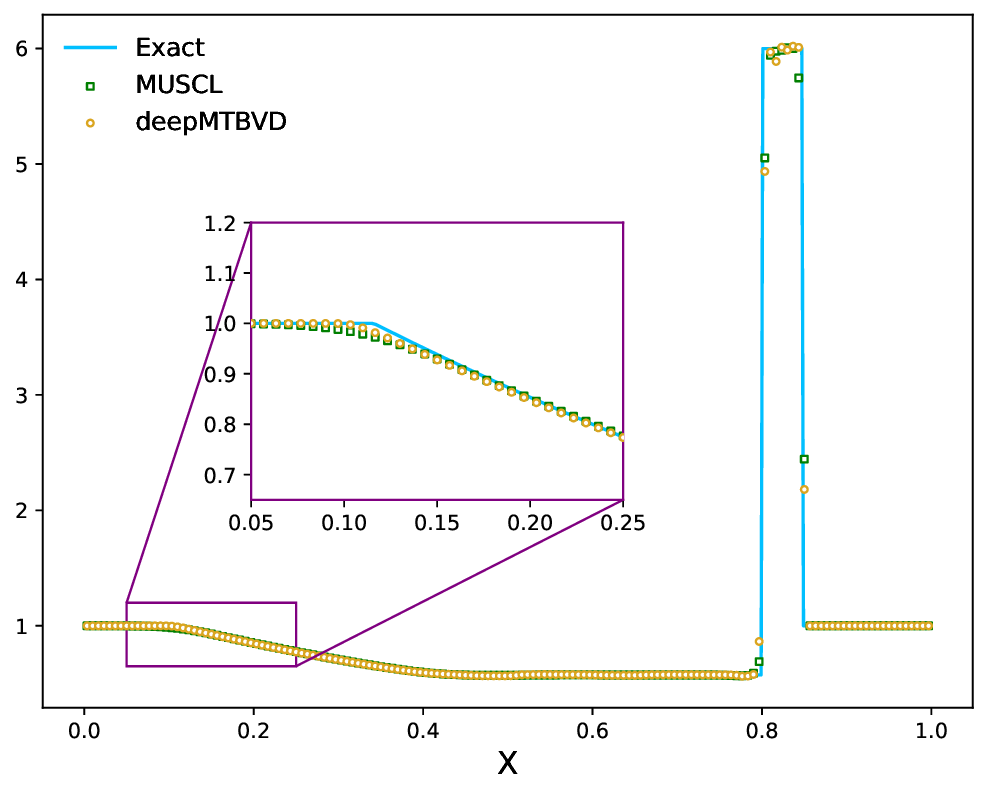}\label{fig:results:case2:a}} 
    \subfigure[Comparison of MUSCL-THINC-BVD and deepMTBVD]{\includegraphics[width=0.49\linewidth]{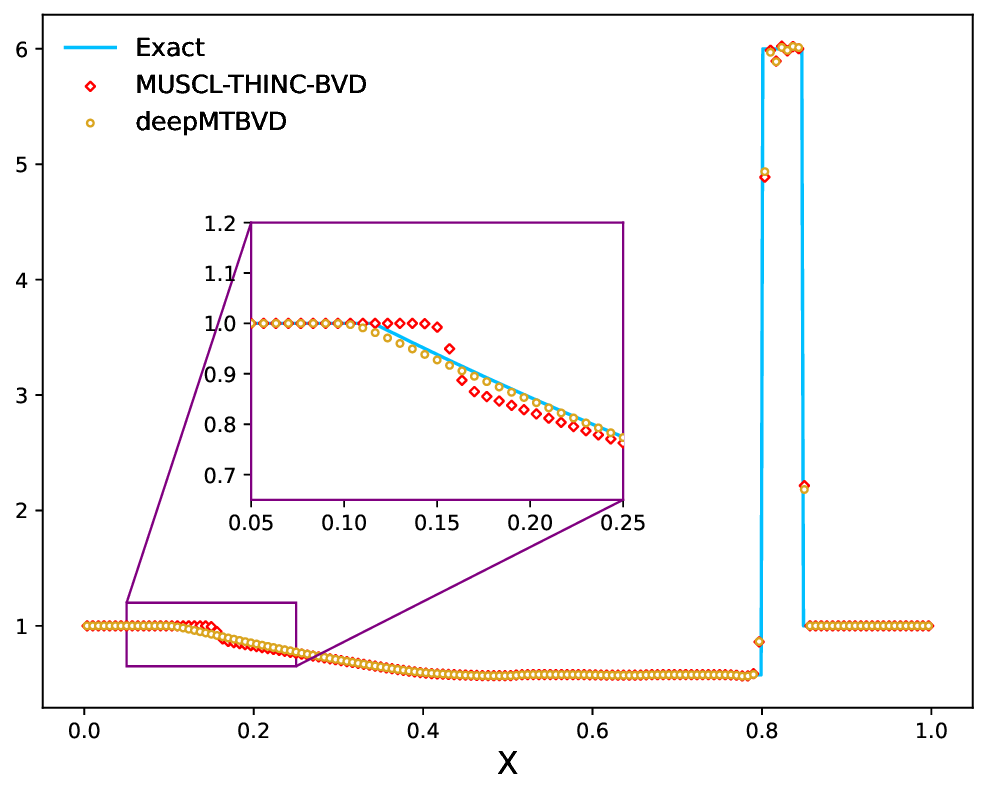}\label{fig:results:case2:b}}
    \caption{ Comparison of numerical results with the exact density solution at $t = 0.012$ with 150 mesh cells. Left: MUSCL (green square) and deepMTBVD (golden circle); Right: MUSCL-THINC-BVD (red diamond) and deepMTBVD (golden circle). The solid line and square denote the exact solution in the instant time.}
    \label{fig:results:case2}
\end{figure}

\begin{figure}[ht!]
    \centering
    \includegraphics[width=\linewidth]{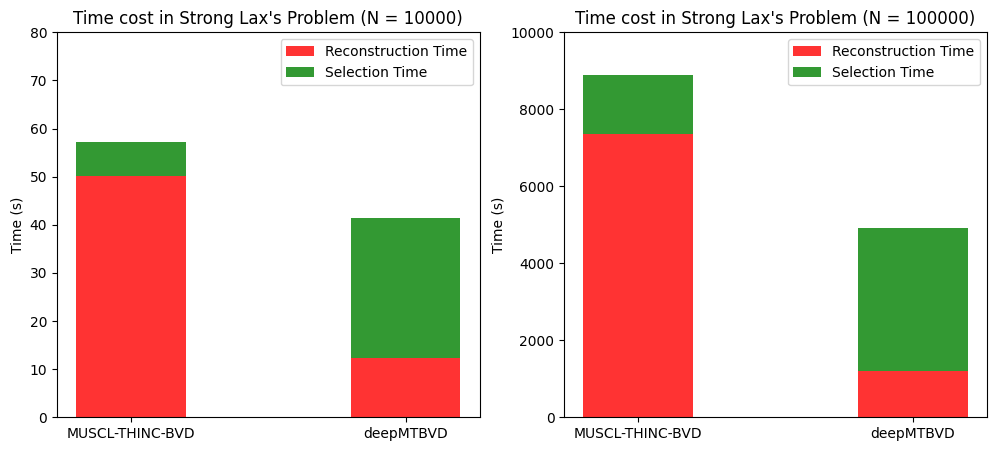}
    \caption{Time cost in Strong Lax's problem. On the left, the mesh size is $N = 10000$ while $N = 100000$ on the right. The deepMTBVD reduces the reconstruction time while it increases the selection time due to the inference by the neural network compared with MUSCL-THINC-BVD. The deepMTBVD scheme takes less time and has higher efficiency and performance than MUSCL-THINC-BVD.}
    \label{fig:result:laxtime}
\end{figure}

The computational domain spans $[0,1]$ and is discretized into 150 uniform cells. The density profiles for both the exact and numerical solutions at time $t = 0.012$ are illustrated in Fig. \ref{fig:results:case2}. As shown in Fig. \ref{fig:results:case2:a}, the deepMTBVD scheme resolves the shock wave more sharply than the MUSCL scheme. Additionally, at the head of the rarefaction wave, the deepMTBVD produces a less dissipative solution compared to the MUSCL scheme. When compared to the MUSCL-THINC-BVD scheme, the deepMTBVD scheme exhibits similar performance in resolving contact discontinuities and shock waves. However, the MUSCL-THINC-BVD scheme tends to generate an overly anti-dissipative solution at the head of the rarefaction wave, while the deepMTBVD maintains a smooth profile, demonstrating its robustness and superiority. Furthermore, Fig. \ref{fig:result:laxtime} highlights the computational time for this example, revealing two key observations: first, although the deepMTBVD scheme increases selection time due to neural network inference, it reduces the overall reconstruction time; second, the deepMTBVD scheme is more time-efficient and produces a higher-quality numerical solution compared to the MUSCL-THINC-BVD scheme.

\begin{figure}[ht!]
  \centering
  \subfigure[Peak overpressure]
  {\includegraphics[width=0.45\textwidth]{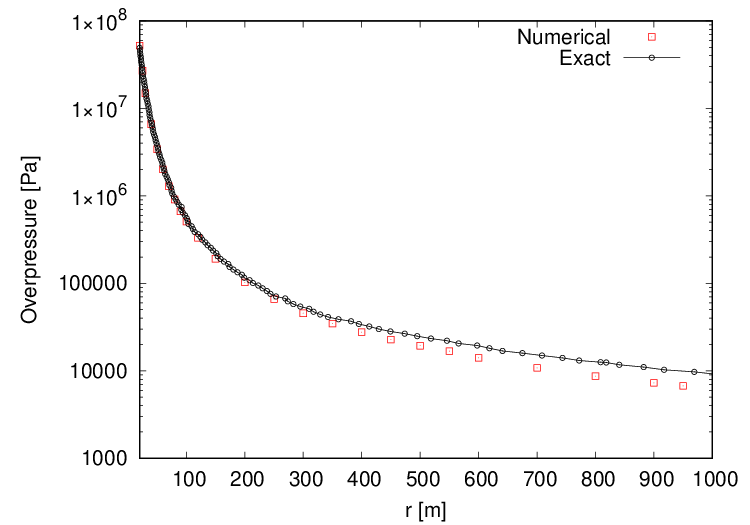}\label{rm:air_blast:a}}
  \subfigure[Impulse]
  {\includegraphics[width=0.45\textwidth]{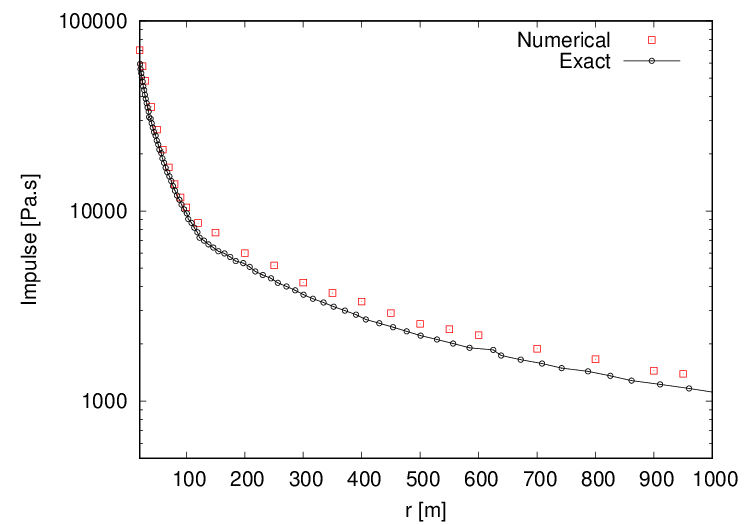}\label{rm:air_blast:b}}\\
  \subfigure[Arrival time]
  {\includegraphics[width=0.45\textwidth]{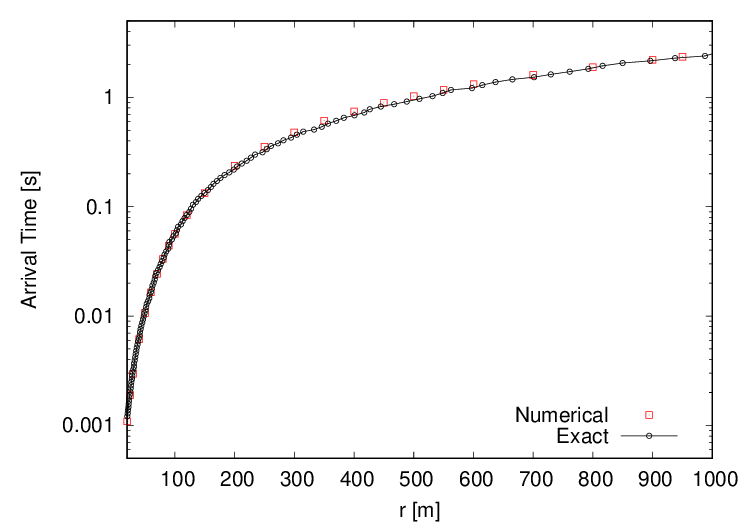}\label{rm:air_blast:c}}
  \subfigure[Positive time duration]
  {\includegraphics[width=0.45\textwidth]{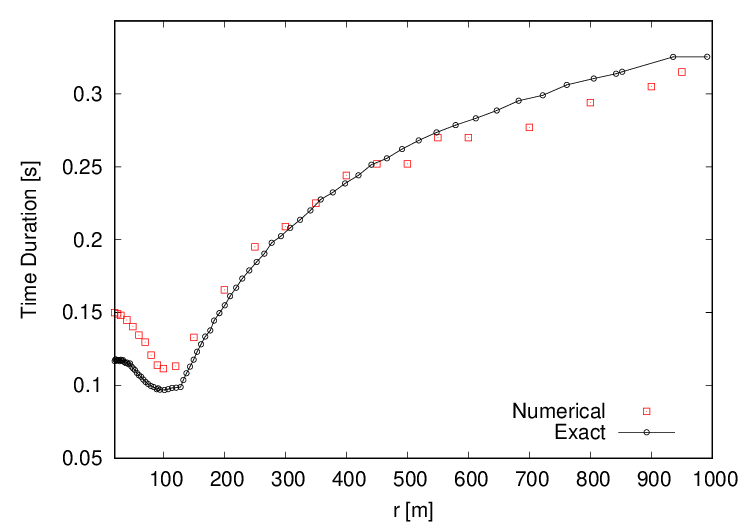}\label{rm:air_blast:d}}
  \caption{Shock wave parameters for air blast problem, including overpressure (upper left), impulse (upper right), arrival time (lower left), positive time duration (lower right). The red square denotes the numerical solution, while the black dot line refers to the exact data.}
  \label{rm::air_blast}
\end{figure}

\subsection{Spherical blast wave problem}
% In this part, we present spherically symmetric blast wave problems, where the governing equations are formulated as follows
% \begin{equation}
% \dfrac{\partial}{\partial t}
% \begin{bmatrix}
% r^2\rho \\ r^2\rho u \\ r^2E
% \end{bmatrix}
% +\dfrac{\partial}{\partial r}
% \begin{bmatrix}
% r^2\rho u \\ r^2(\rho u^2+p) \\ r^2(E+p)u
% \end{bmatrix}
% =\begin{bmatrix}
% 0 \\ 2rp \\ 0
% \end{bmatrix}.
% \label{eq:spheq}
% \end{equation}
% The source term in \eqref{eq:spheq} is discretized using an explicit Euler method. 
In this part, we present spherically symmetric blast wave problems. The governing equations are reformulated by the spherical coordinates system. The extensive blast wave that propagates through the air due to the nuclear explosion is commonly referred to as the \emph{blast wave}. In this example, we simulate the blast wave from one kiloton of nuclear charge. The explosion products and air are modeled by the ideal gas EOS with adiabatic exponents $\gamma=1.2$ and $1.4$, respectively. The initial density and pressure are $618.935 \text{ kg}/\text{m}^3$ and $6.314 \times 10^{12}\text{ Pa}$ for the explosion products, and $1.29\text{ kg}/\text{m}^3$ and $1.013\times 10^5\text{ Pa}$ for the air. The phase interface is initially located at $r=0.3\text{ m}$. To effectively capture the wave propagation, we use a computational domain of radius $5000\text{ m}$.

The diagram in Fig. \ref{rm::air_blast} illustrates four physical quantities, which describe the effect and influence of the blast wave. It is known that the destructive effects of the blast wave can be measured by its \emph{overpressure}, i.e., the amount by which the static pressure in the blast wave exceeds the ambient pressure ($1.013\times 10^5\text{ Pa}$). The overpressure increases rapidly to a peak value when the blast wave arrives, followed by a roughly exponential decay. The integration of the overpressure over time is called \emph{impulse}. The time when the shock wave arrives is called \emph{arrival time}. The duration of pressure larger than atmospheric pressure is called \emph{positive time duration}. Fig. \ref{rm:air_blast:a} to Fig. \ref{rm:air_blast:d} show the peak overpressure, impulse, shock arrival time at different radii, and positive time duration. The results are compared with the point explosion solution in \cite{qiao2003}, confirming the accuracy of the methods in the air blast applications.

\subsection{Two-dimensional problems}
In this part, we present several two-dimensional problems in engineering applications carried out on the Cartesian grid for each phase of fluids. In these problems, parallel computing based on the classical domain decomposition methods is implemented to improve the efficiency of the simulation.

\subsubsection{Double Mach reflection problem}
This example simulates a 10 Mach propagating planar shock with the 30-degree ramp originally proposed in \cite{woodward1984}. This test is utilized to mimic the intense reflected and refracted shock waves, along with the more elaborate vortical structures, which are particularly responsive to the dissipation characteristics of numerical algorithms. The efficacy of a numerical scheme in reducing dissipation is assessed by the vortical structures that arise from Kelvin-Helmholtz instabilities along the slip line within the recirculation zone. Schemes characterized by significant numerical dissipation tend to yield less intricate structural details.

The whole computational region is $[0, 4] \times [0, 1]$. The reflecting wall, beginning at $x = \frac{1}{6}$, lies along the bottom boundary and the short region from $x \in [0, \frac{1}{6}]$ along the bottom boundary. A right-going shock is imposed with $60$ degrees relative to the $x$-axis and extends to the top $y = 1$. The left boundary and the short region from $x \in [0, \frac{1}{6}]$ on the bottom are assumed to be the initial post-shock flow. The right boundary condition is given by zero-gradient. The solutions are computed at time $t = 0.2$ with uniform grid mesh $\Delta x = \Delta y = \frac{1}{160}$. We set $\kappa_{ref} = 0.4$ in this case.

\begin{figure}[ht!]
    \centering
    \subfigure[MUSCL]{\includegraphics[width=0.9\linewidth]{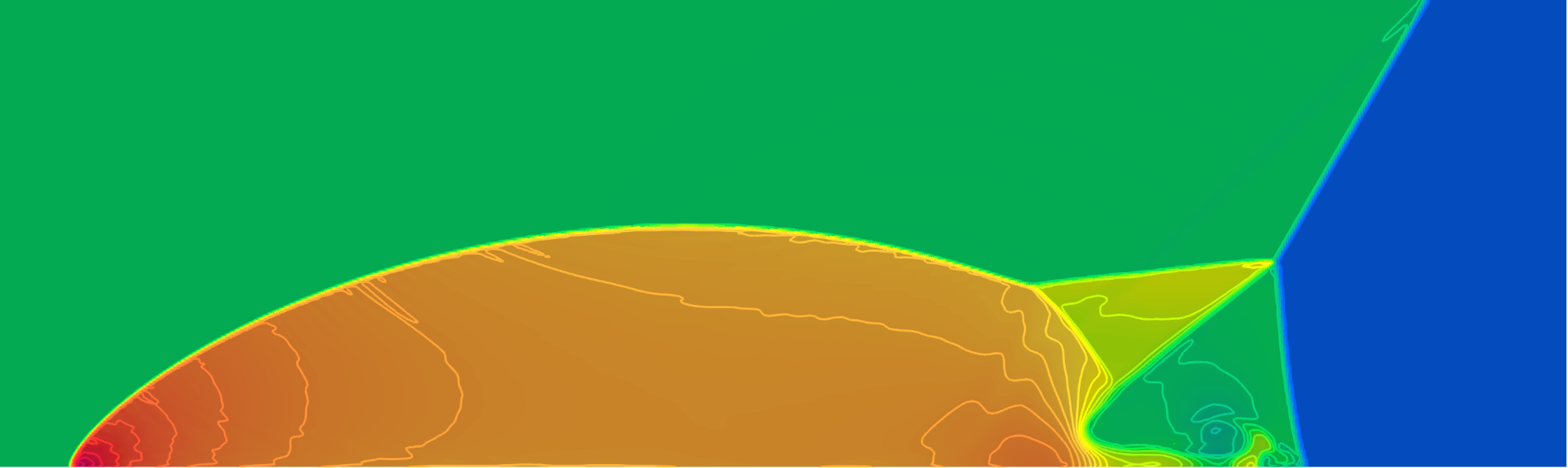}} \\
    \subfigure[deepMTBVD]{\includegraphics[width=0.9\linewidth]{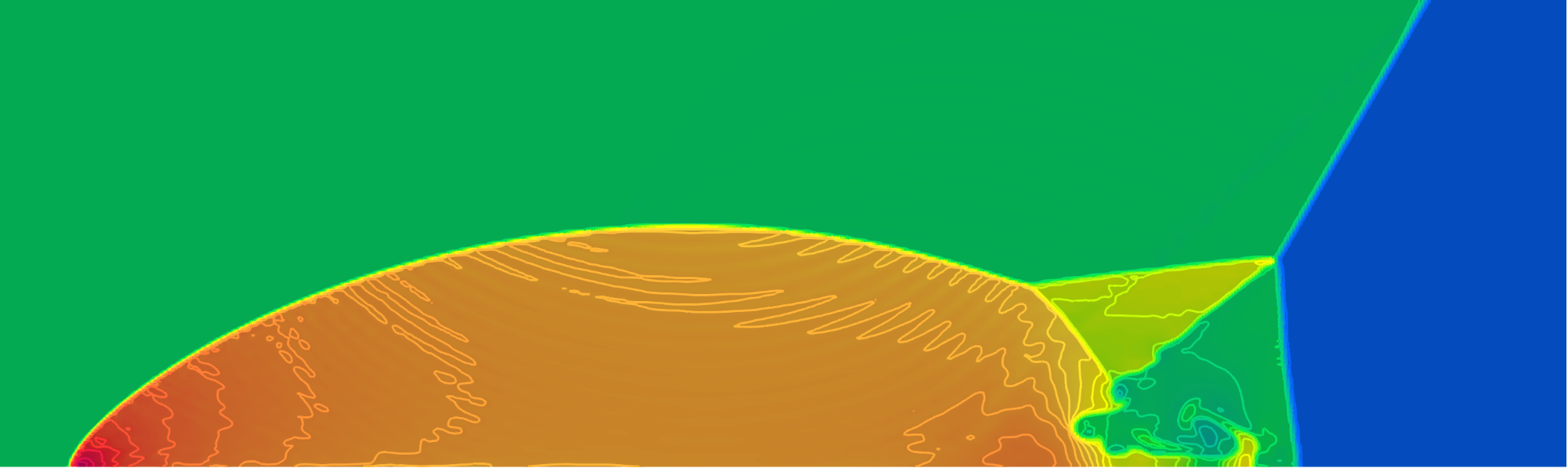}} \\
    \subfigure[MUSCL-THINC-BVD]{\includegraphics[width=0.9\linewidth]{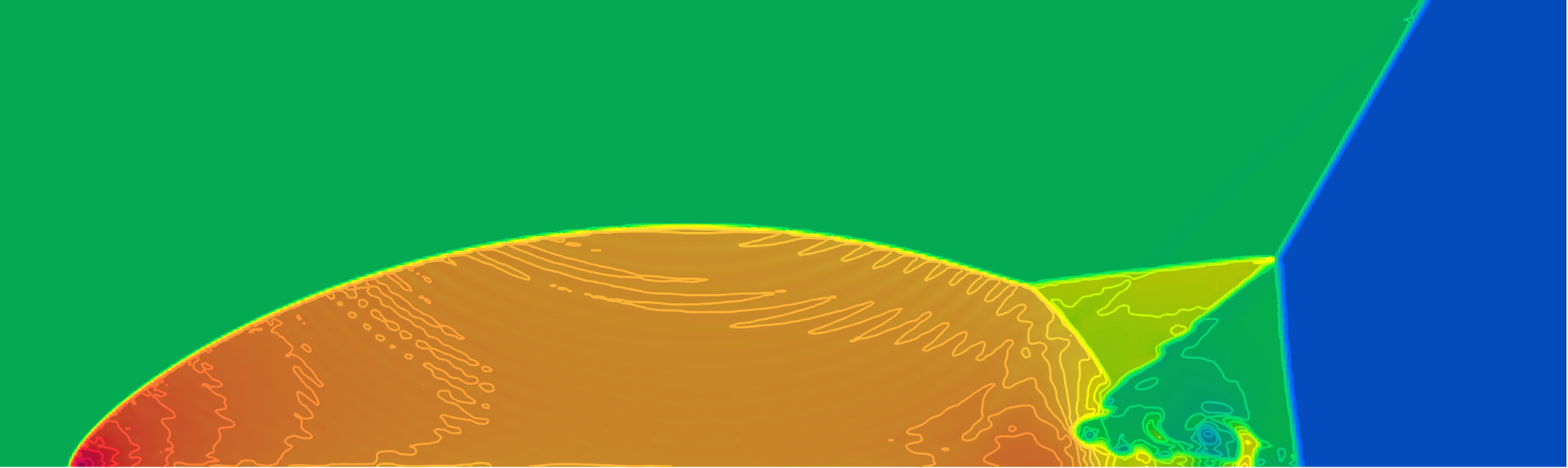}}
    \caption{Density contours range from 1.5 to 21.5 with 30 levels at time $t = 0.2$.}
    \label{fig:result:doulemach}
\end{figure}

%We present the numerical results with 30 equidistant levels ranging from 1.5 to 21.5 in Fig. \ref{fig:result:doulemach}. From the results, it is observed that the deepMTBVD scheme is able to capture the shock without spurious oscillation. Compared to the MUSCL scheme, the MUSCL-THINC-BVD and deepMTBVD schemes show less dissipation and resolve fine vertical structures along the slip line.

The numerical results, depicted in Fig. \ref{fig:result:doulemach} with 30 equidistant contour levels ranging from 1.5 to 21.5, demonstrate that the deepMTBVD scheme effectively captures the shock without introducing spurious oscillations. In comparison to the MUSCL scheme, both the MUSCL-THINC-BVD and deepMTBVD schemes exhibit lower numerical dissipation and better resolve fine vortical structures along the slip line.

\subsubsection{Nuclear air blast near rigid ground}
\begin{figure}[ht!]
\centering
\subfigure[Peak overpressure]
{\includegraphics[width=0.45\textwidth]{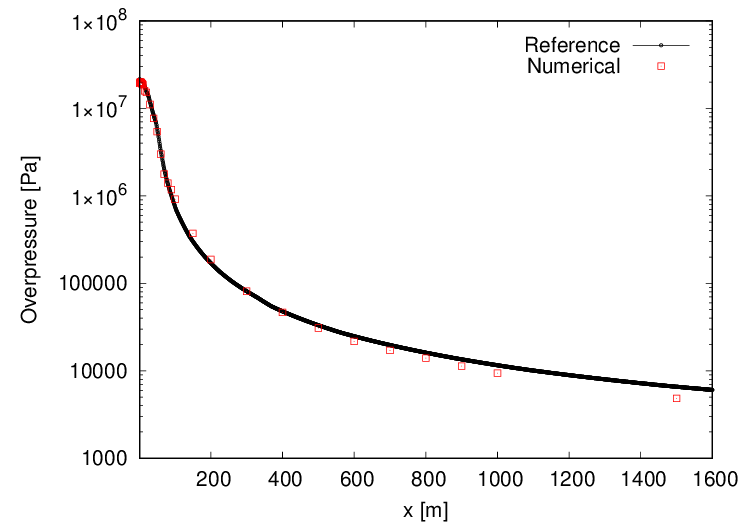}} 
\subfigure[Impulse]
{\includegraphics[width=0.45\textwidth]{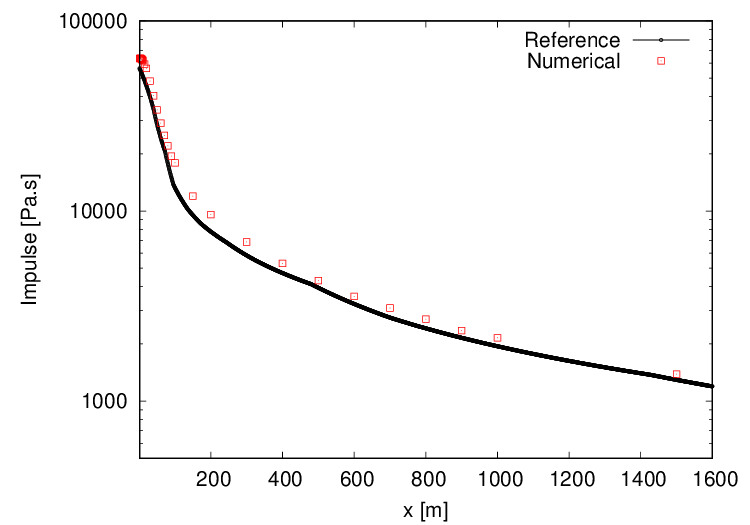}} 
\caption{Blast wave parameters of typical radii on the ground.}
\label{res:blast_paras}
\end{figure}

\begin{figure}[ht!]
\centering
\subfigure[$t=0.002$s]
{\includegraphics[width=0.8\textwidth]{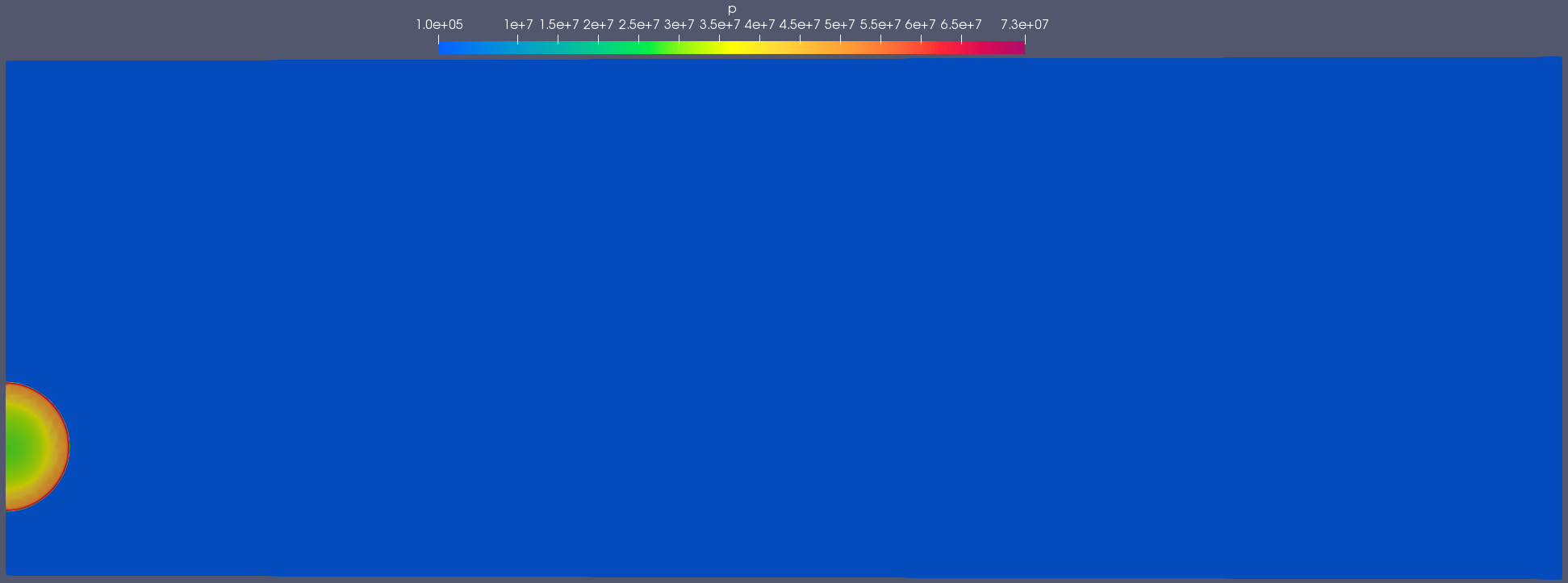}}     \\
% \vspace{-2mm}
\subfigure[$t=0.04$s] 
{\includegraphics[width=0.8\textwidth] {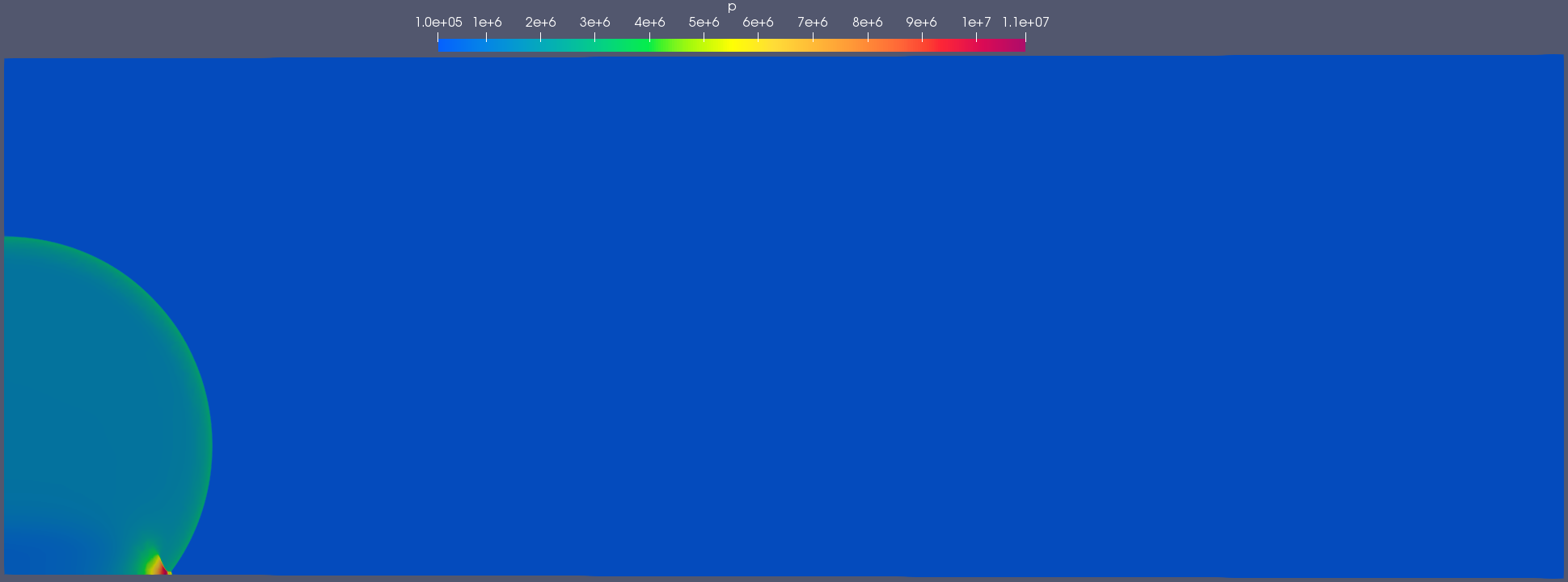}}   \\
% \vspace{-2mm}
\subfigure[$t=0.1$s] 
{\includegraphics[width=0.8\textwidth] {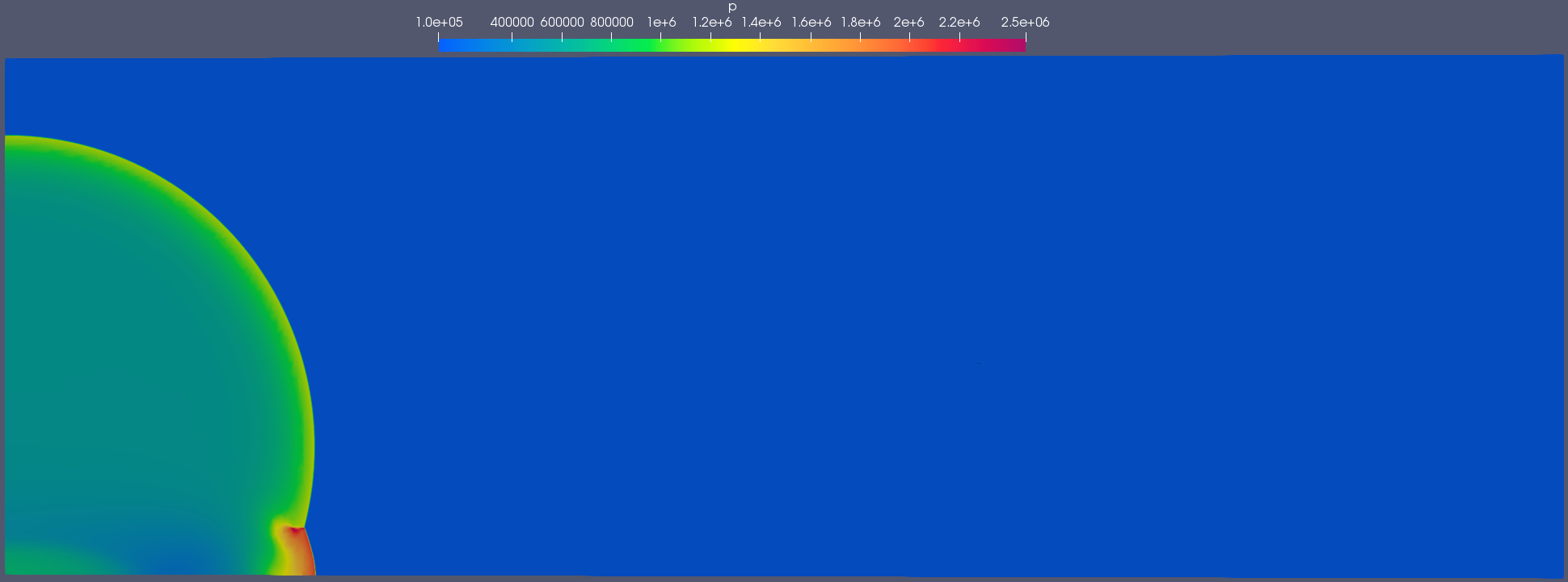}}  \\
% \vspace{-2mm}
\caption{Pressure contours of air blast problem at the typical time.}
\label{res:airblast:0}
\end{figure}

\begin{figure}[htbp]
\centering
\subfigure[$t=0.3$s] 
{\includegraphics[width=0.8\textwidth] {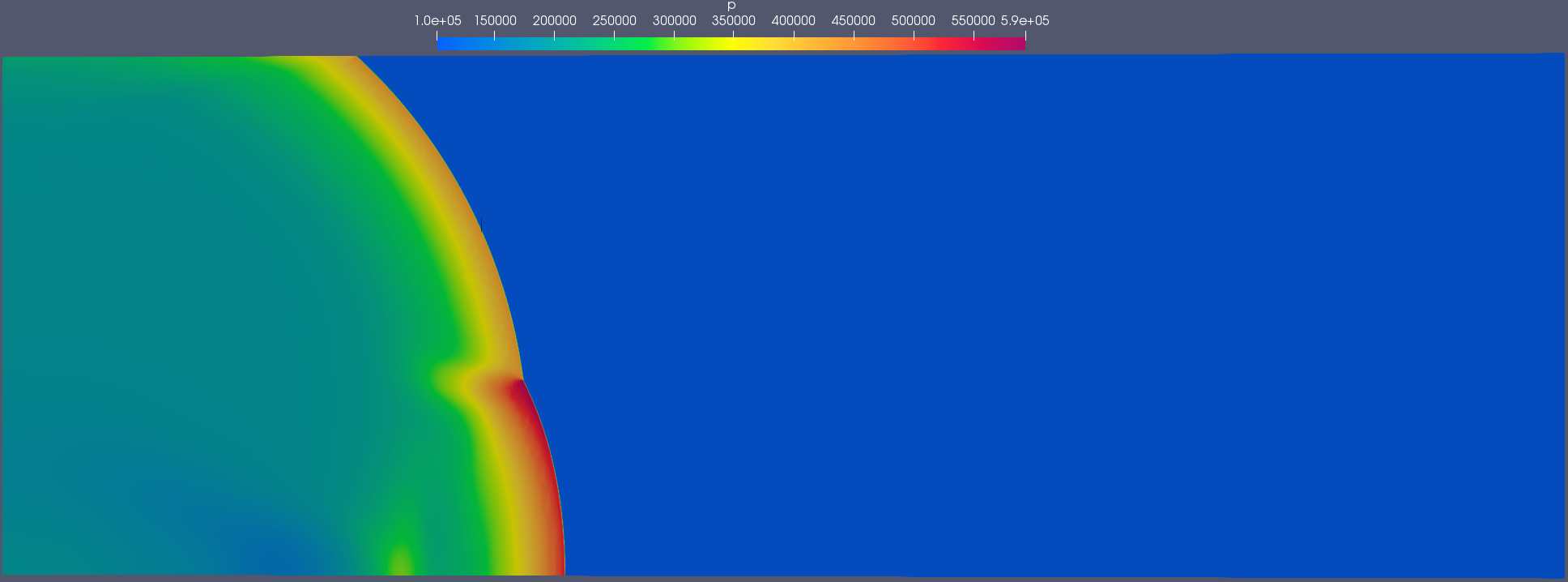}}\\
\subfigure[$t=0.7$s]
{\includegraphics[width=0.8\textwidth] {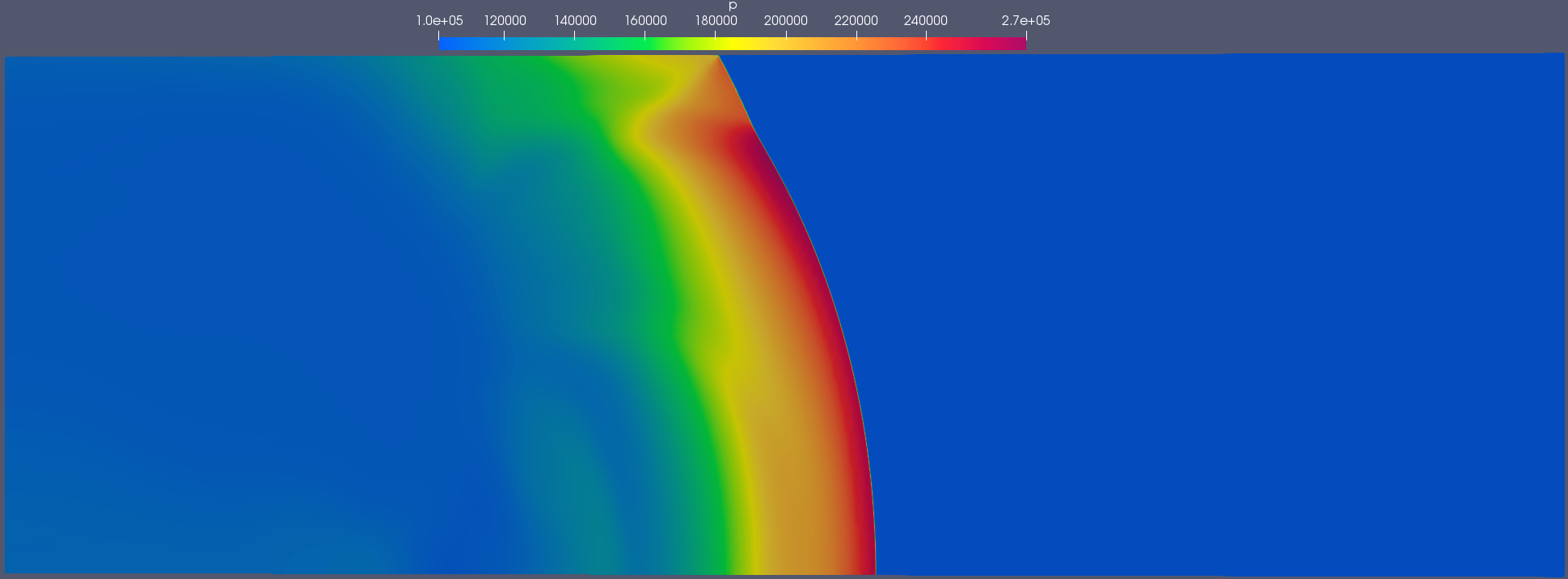}}  
\caption{Pressure contours of air blast problem at the typical time (continued).} 
\label{res:airblast}
\end{figure}

% In this example, we simulate a nuclear air blast problem within the cylindrically symmetric coordinate system, which has the following governing equations
% \[
% \dfrac{\partial}{\partial t}
% \begin{bmatrix}
% r\rho \\ r\rho u \\ r\rho v\\ rE
% \end{bmatrix}
% +\dfrac{\partial}{\partial r}
% \begin{bmatrix}
% r\rho u \\ r(\rho u^2+p) \\ r\rho uv \\ r(E+p)u
% \end{bmatrix}
% +\dfrac{\partial}{\partial z}
% \begin{bmatrix}
% r\rho v \\ r\rho uv \\ r(\rho v^2+p)\\ r(E+p)v
% \end{bmatrix}
% =\begin{bmatrix}
% 0 \\ p \\ 0 \\ 0
% \end{bmatrix}.
% \]
In this example, we simulate a nuclear air blast problem within the cylindrically symmetric coordinate system. The computational domain is $(0,400)\times(0,1500)$ in meters, and the burst point is at $(0, 50)$m with an initial radius of 0.3m. The governing equation of state of this problem is expressed in a cylindrical form \cite{Chen2018} and the initial conditions are 
\[
[\rho, u, v, p, \alpha_{nuc}]^\top = \left\{
\begin{array}{ll}
[618.935, ~0, ~0, ~6.314 \times 10^{12}, 1-10^{-8}]^\top, & \sqrt{x^2+(y-50)^2} \le 0.3,\\ [1mm]
[1.29, ~0, ~0, ~1.013\times 10^5, 10^{-8}]^\top, & \sqrt{x^2+(y-50)^2} > 0.3,
\end{array}
\right.
\]
where $\alpha_{nuc}$ is the volume fraction of explosion products. The explosion products and air are modeled by the ideal gas EOS with adiabatic exponents $\gamma=1.2$ and $1.4$, respectively. All of the boundaries are set as outflow conditions except that the bottom edge $y=0$ is a rigid ground. Fig. \ref{res:airblast:0} and Fig. \ref{res:airblast} illustrate the pressure contours at a typical time. When the blast wave produced by the nuclear explosion arrives at the rigid ground, it will be reflected first and propagate along the rigid ground simultaneously. When the incident angle exceeds the limit, the reflective wave switches from regular to irregular, and a Mach blast wave occurs. The peak overpressure and impulse at different radii are shown in Fig. \ref{res:blast_paras}, and agree well with the reference data interpolated from the experimental data in \cite{Glasstone1977}.

\subsection{Three-dimensional problems}
In this part, we present several three-dimensional problems in engineering applications carried out on the Cartesian grid for each phase of fluids. Similar to two-dimensional calculations, parallel computing based on the classical domain decomposition methods is implemented to improve the efficiency of the simulation.

\subsubsection{TNT explosion in a local street environment}
\begin{figure}[htbp]
\centering
\includegraphics[width=0.9\textwidth]{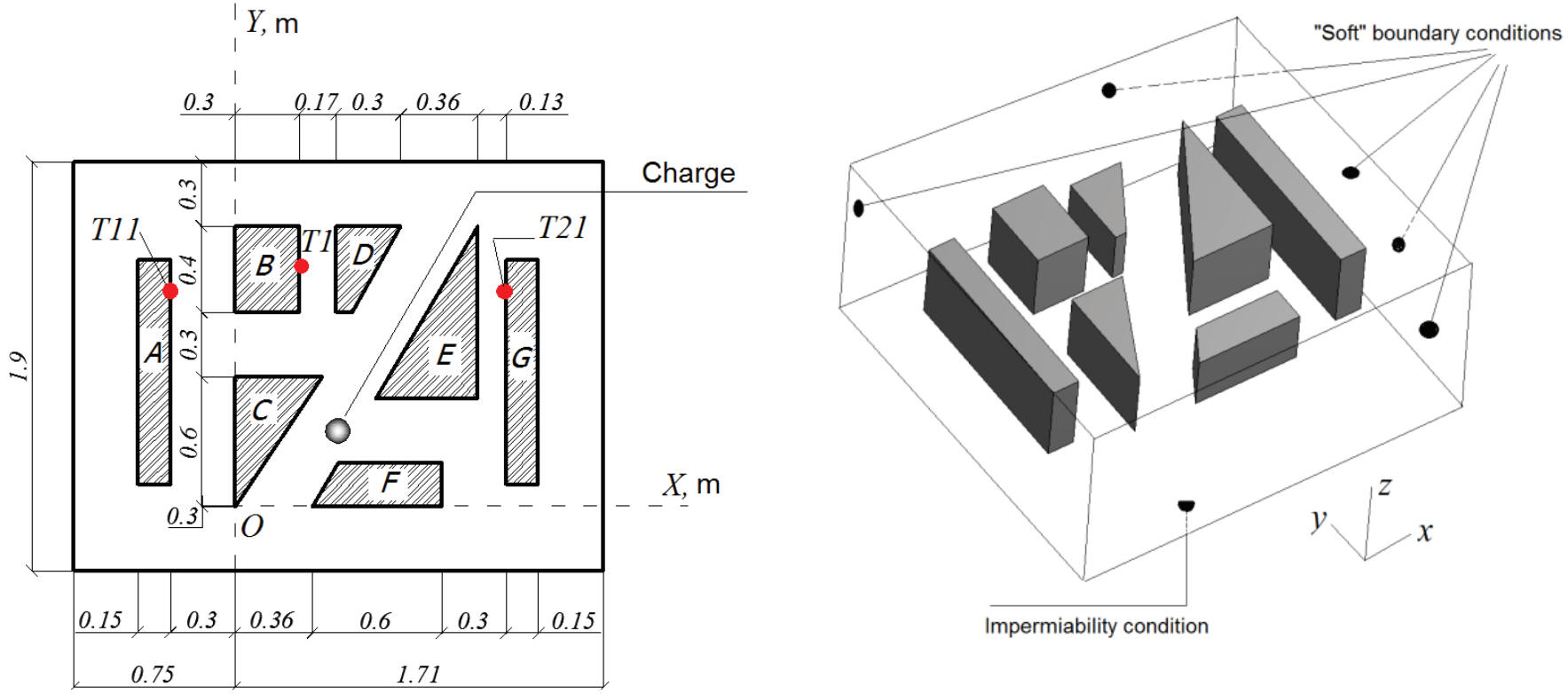}
\caption{Experimental model of TNT explosion in local urban scaled building group. Right: the location of the gauge points, T1 and T2, shown as red circles. Left: the boundary condition of the computational domain.}
\label{airblast3d:local_model}
\end{figure}

In this example, we calculate a shock wave propagation process within a local urban scaled building group, as described in \cite{Fedorova2016Simulation}. The experiment employed a spherical charge, with an equivalent TNT equivalent of 0.16 kg, at a height of 0.04 m. The experimental model is shown in Fig.~\ref{airblast3d:local_model}, where the ground and building scaling models have sufficient stiffness, and the deformation during the shock wave is neglected. Three measurement points are distributed on the building surface to obtain the pressure-time history of the flow field. 

A three-dimensional numerical simulation was conducted for this experimental scenario, with the building exterior and ground being treated as structural boundaries and the other settings being treated as non-reflection boundaries. The model used a structured grid with a global grid size of 0.5 cm. 
\begin{figure}[ht!]
\centering
\subfigure[$t=0.2$ms]
{\includegraphics[width=0.6\textwidth]{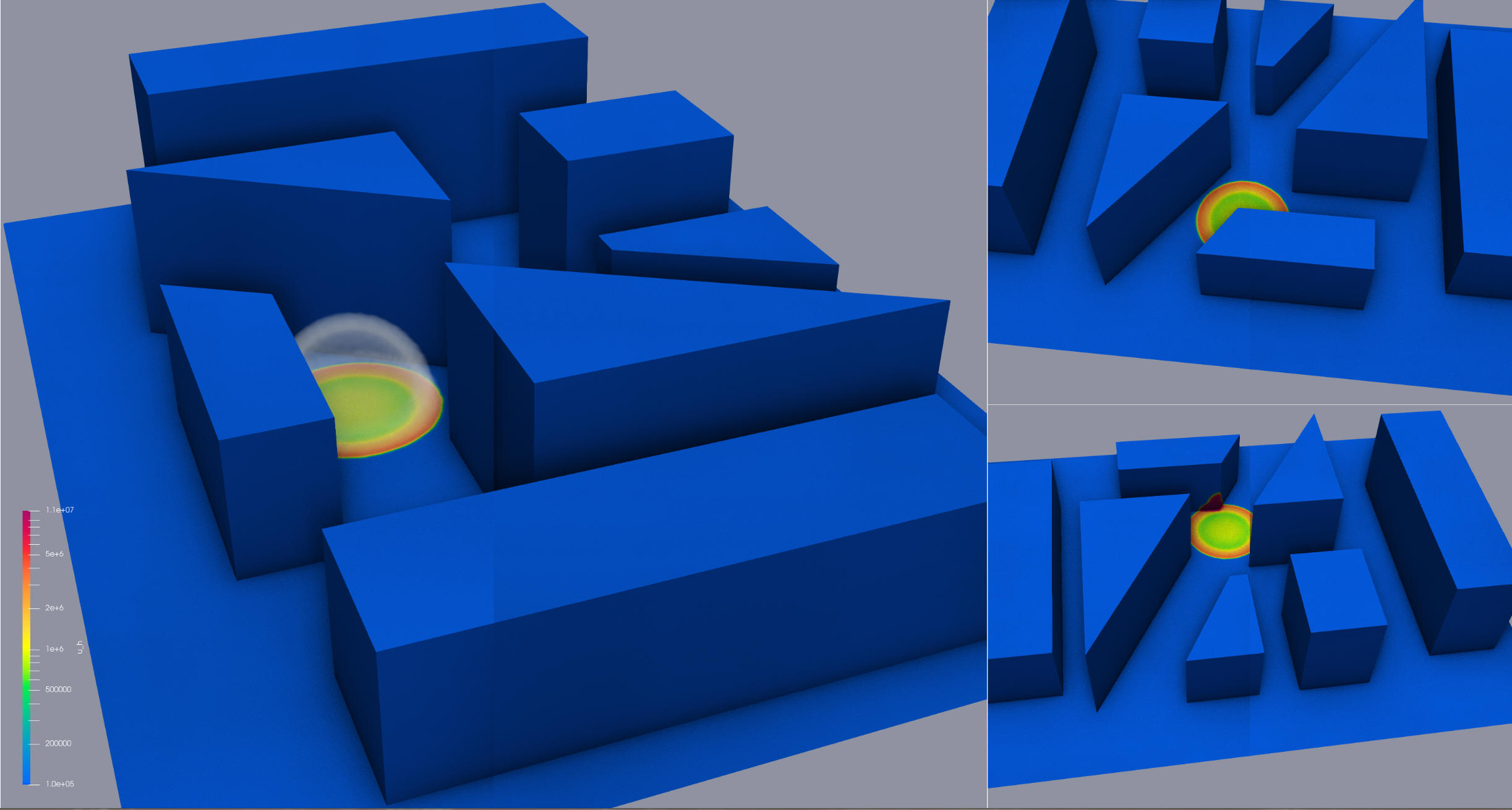}}  \\
% \vspace{-2mm}
\subfigure[$t=0.6$ms]
{\includegraphics[width=0.6\textwidth]{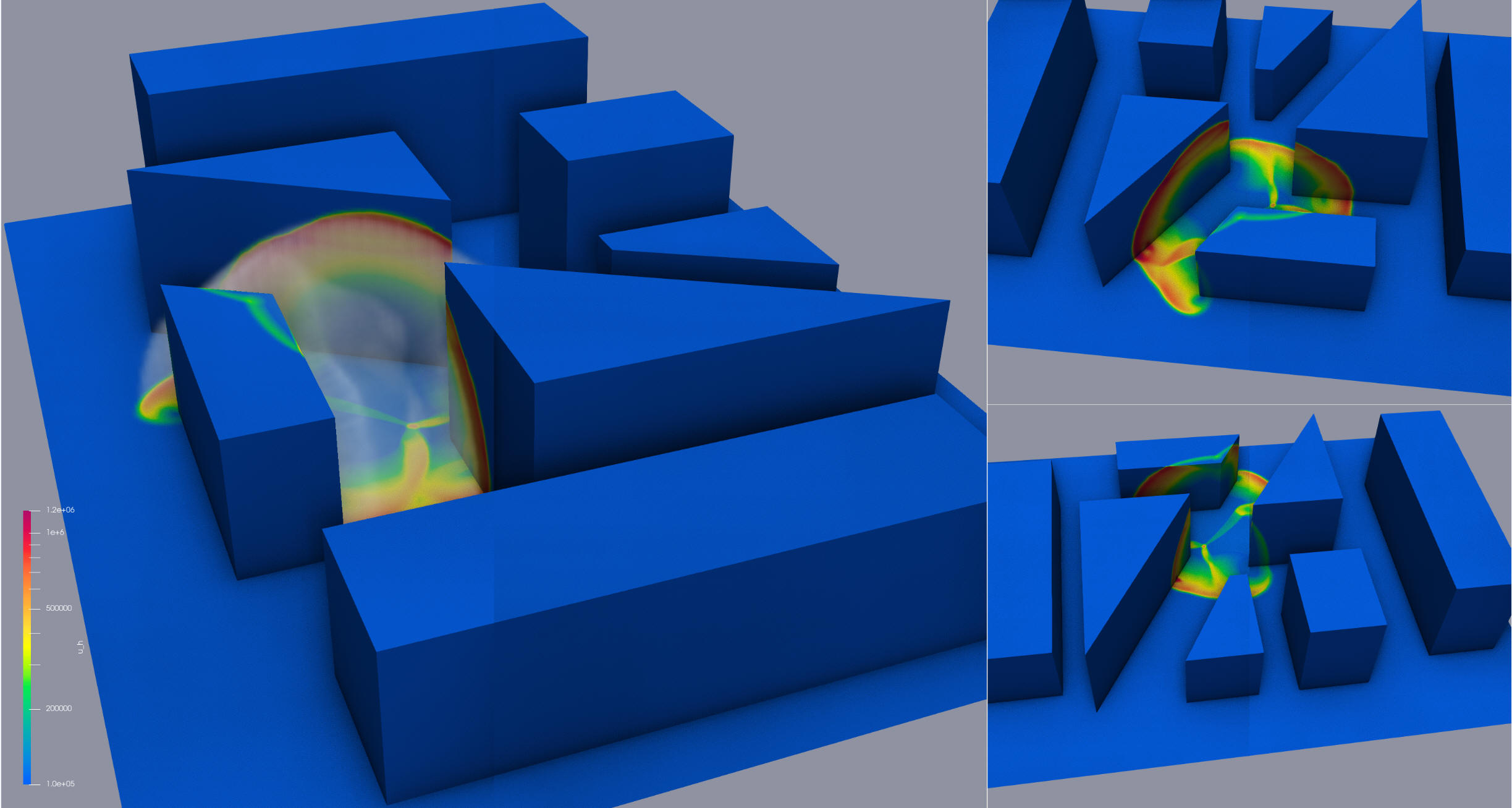}}  \\
% \vspace{-2mm}
\subfigure[$t=1.2$ms]
{\includegraphics[width=0.6\textwidth]{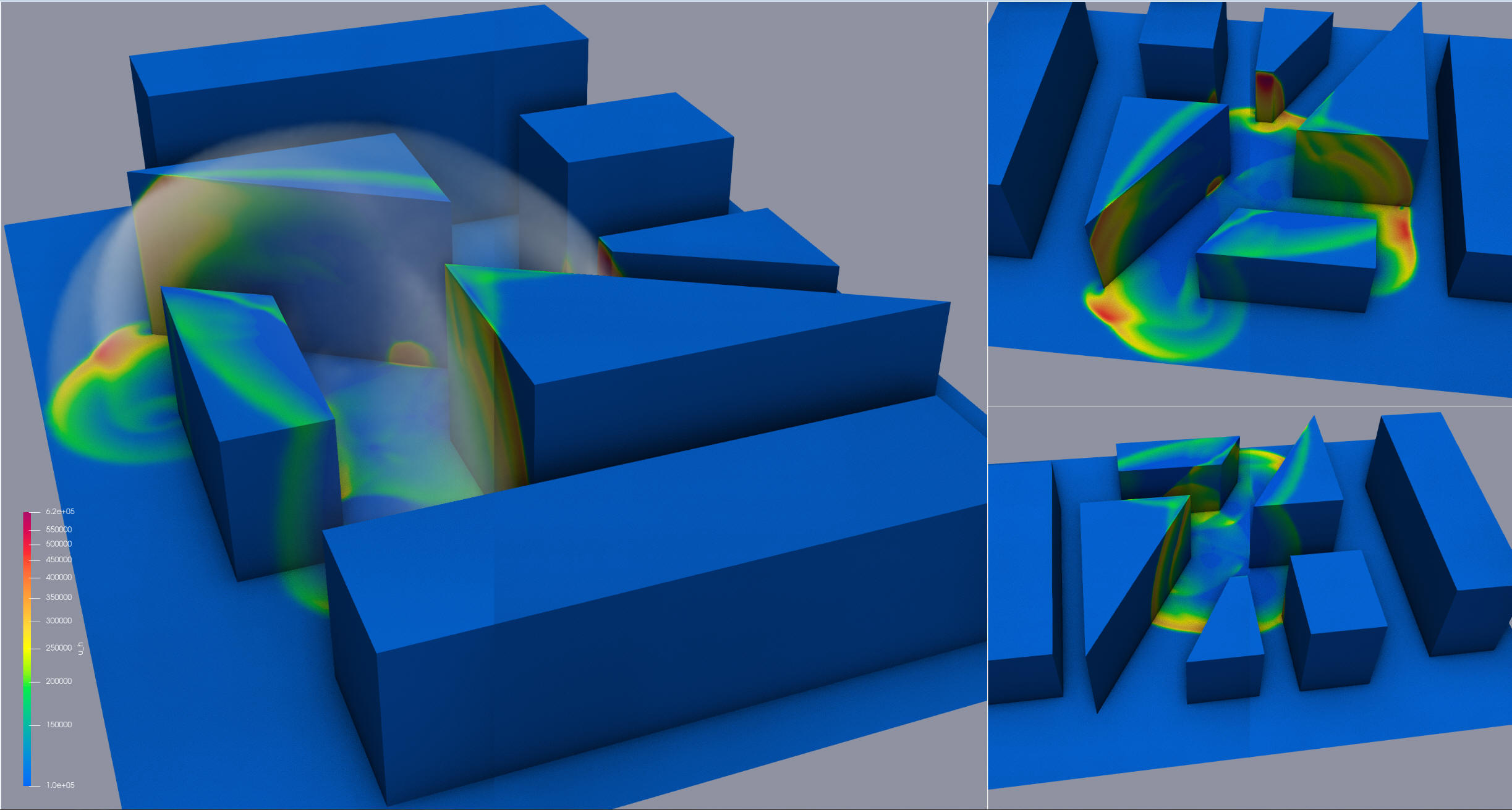}}
\caption{The propagation of shock waves within a local urban building group.}
\label{airblast3d:local}
\end{figure}

\begin{figure}[ht!]
\centering
\subfigure[The time history of the pressure at the gauge point T1]
{\includegraphics[width=0.48\textwidth]{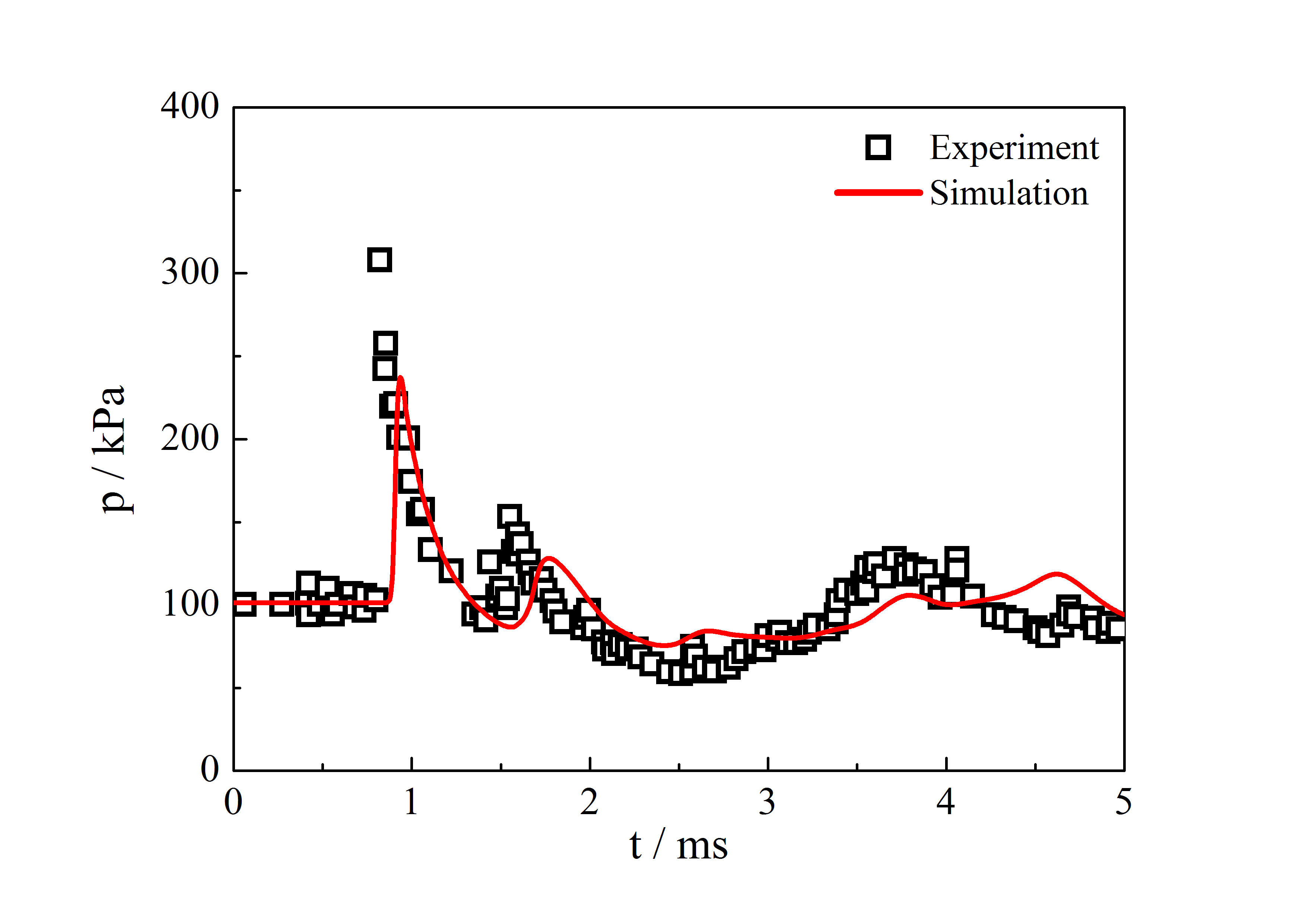}}
\subfigure[The time history of the pressure at the gauge point T2]
{\includegraphics[width=0.48\textwidth]{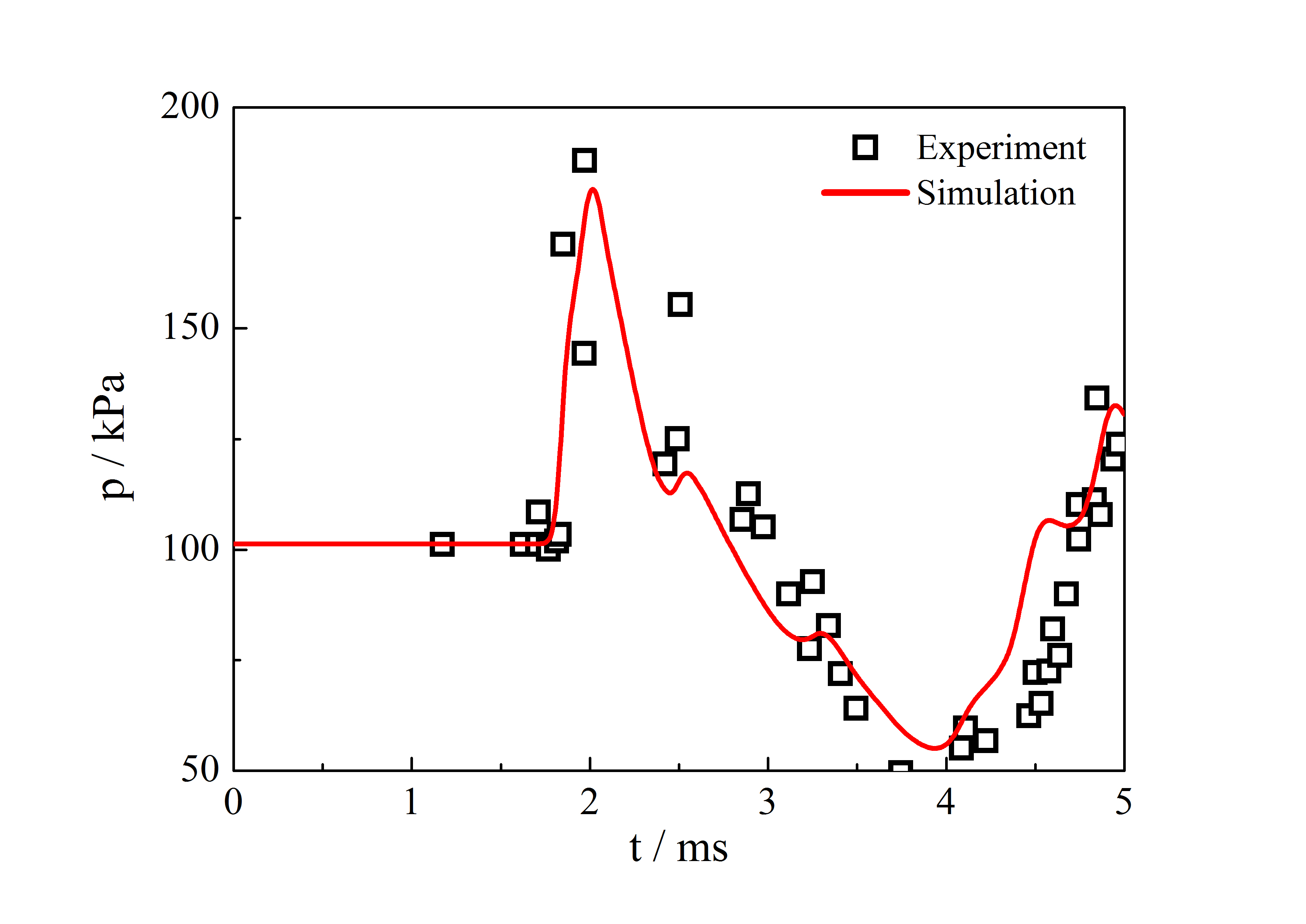}}
\caption{The time history of shock waves at the gauge points in the experimental setup. The gauge points, T1 and T2, are plotted as red circles in Fig \ref{airblast3d:local_model}. The history of the resulting pressure is in great agreement with the experimental data in \cite{Fedorova2016Simulation}. }
\label{airblast3d:local_para}
\end{figure}

Fig.~\ref{airblast3d:local} presents a typical instantaneous pressure cloud diagram of the numerical simulation, indicating that the shock wave interacts with different buildings sequentially in the propagation process, forming a complex wave series structure within the flow field. The numerical simulation results can accurately reflect the propagation process of shock waves. Compared to the experimental data from \cite{Fedorova2016Simulation} obtained in T1 and T2 (illustrated as red circles in Fig. \ref{airblast3d:local_model}), the numerical simulation in Fig. \ref{airblast3d:local_para} demonstrates excellent agreement with the resulting pressure at the gauge point.

\section{Conclusion} \label{sec:conclusion}
We propose a hybrid scheme, combining the multi-component diffuse interface method with the material point method on structured Eulerian grids, to simulate the complex multiphase flow under extreme conditions in large-scale urban environments. A seven-equation model with an arbitrary number of phases is adopted to simulate the multiphase compressible flows, and the general Godunov method is applied to calculate the numerical fluxes of the conservative and non-conservative terms. The material point method with the rigid solid constitutive model is used to simulate the urban buildings and irregular ground. An artificial neural network equation of state is proposed to simulate the intensive explosion products and real gas under extreme pressure and temperature. A new deepMTBVD reconstruction scheme from our previous work is extended to the multiphysics system and holds properties of high resolution and high efficiency. With the addition of high-performance parallel computation, a robust and efficient multi-physical numerical system is established to solve the intense blast wave problems in large-scale and complex urban environments. Finally, several benchmark examples and large-scale air blast problems in engineering applications are simulated to validate the schemes.

\section{Acknowledgments}
% This work is supported by the Fundamental Research Funds for the Central Universities. It is also partially supported by the National Key R$\&$D Program of China (Project No. 2020YFA0712000), the Shanghai Science and Technology Innovation Action Plan in Basic Research Area (Project No. 22JC1401700), the Strategic Priority Research Program of Chinese Academy of Sciences (Grant No. XDA25010405) and the National Natural Science Foundation of China (Grant No. DMS-11771290).  Lidong Cheng's work is supported by the Postdoctoral Fellowship Program of CPSF (Grant No. GZC20231592).
It is partially supported by the Shanghai Science and Technology Innovation Action Plan in Basic Research Area (Project No. 22JC1401700) and the National Natural Science Foundation of China (Grant No. DMS-11771290). Lidong Cheng's work is supported by the Postdoctoral Fellowship Program of CPSF (Grant No. GZC20231592). This work is also supported by the Fundamental Research Funds for the Central Universities.

%%%% Bibliography  %%%%%%%%%%
\bibliographystyle{unsrt}
% \biboptions{sort&compress} %used for [1-4] not [1,2,3,4]
\bibliography{ref}
\end{document}